\documentclass{article}

\usepackage{arxiv}

\usepackage[compact]{titlesec}
\usepackage[utf8]{inputenc}
\usepackage[T1]{fontenc}    
\usepackage{hyperref}       
\usepackage{amsfonts}       
\usepackage{nicefrac}       
\usepackage{microtype}      
\usepackage{amsmath}
\usepackage{cleveref}       
\usepackage{lipsum}         
\usepackage{wrapfig}
\usepackage{xcolor}
\usepackage{subfigure}
\usepackage{url}
\usepackage{diagbox}
\usepackage[title]{appendix}
\usepackage{graphicx}
\usepackage{tikz}
\usepackage{booktabs}
\usepackage{enumitem}
\usepackage{multirow}
\usepackage{algorithm} 
\usepackage{algorithmicx}
\usepackage{algpseudocode}
\usepackage{bbm}
\usepackage{caption}
\usepackage{filecontents}
\usepackage{ragged2e}

\newcommand{\kj}[1]{{#1}} 

\newcommand{\zhaoyuan}[1]{{#1}}
\newcommand{\yifan}[1]{{#1}}

\newenvironment{sitemize}{%
        \begin{list}{$\bullet$}{%
                        \setlength{\itemsep}{2pt}%
                        \setlength{\leftmargin}{1.2em}%
                        \setlength{\topsep}{2pt plus 2pt minus 2pt}%
                        \setlength{\parsep}{0.0cm}}%
        }{\end{list}}

\long\def\comment#1{}
\newcommand{\para}[1]{\smallskip\noindent {\bf #1}}
\newcommand{\ie}{{\em i.e.}}
\newcommand{\eg}{{\em e.g.}}

\usepackage[switch]{lineno}

\title{Real-Time Neural Video Recovery and Enhancement on Mobile Devices}

\date{}

\author{Zhaoyuan He\textsuperscript{1},~~~~
Yifan Yang\textsuperscript{2},~~~~
Lili Qiu\textsuperscript{1,2},~~~~
Kyoungjun Park\textsuperscript{1}\\
\textsuperscript{1}Department of Computer Science, The University of Texas at Austin, Austin, TX, USA\\
\textsuperscript{2}Microsoft Research Asia Shanghai, Shanghai, P.R.China}

\renewcommand{\headeright}{}
\renewcommand{\undertitle}{}

\hypersetup{
pdftitle={A template for the arxiv style},
pdfsubject={q-bio.NC, q-bio.QM},
pdfauthor={David S.~Hippocampus, Elias D.~Striatum},
pdfkeywords={First keyword, Second keyword, More},
}

\begin{document}
\maketitle

\begin{abstract}
As mobile devices become increasingly popular for video streaming, it's crucial to optimize the streaming experience for these devices. Although deep learning-based video enhancement techniques are gaining attention, most of them cannot support real-time enhancement on mobile devices. Additionally, many of these techniques are focused solely on super-resolution and cannot handle partial or complete loss or corruption of video frames, which is common on the Internet and wireless networks.

To overcome these challenges, we present a novel approach in this paper. Our approach consists of (i) a novel video frame recovery scheme, (ii) a new super-resolution algorithm, and (iii) a receiver enhancement-aware video bit rate adaptation algorithm. We have implemented our approach on an iPhone 12,
and it can support 30 frames per second (FPS). We have evaluated our approach in various networks such as WiFi, 3G, 4G, and 5G networks. Our evaluation shows that our approach enables real-time enhancement and results in a significant increase in video QoE (Quality of Experience) of 24\% - 82\% in our video streaming system.

\end{abstract}

\section{Introduction}
\label{sec:intro}

\para{Motivation:} Machine learning (ML) has seen tremendous progress and penetrated almost every aspect of our life. One of the major applications of ML is to apply it to enhancing video quality. In particular, videos account for the majority of Internet traffic. However, the Internet bandwidth is highly fluctuating and hard to predict; when the network condition degrades, the video can get lost or its sending rate has to decrease. Many ML-based video enhancement approaches have been developed.

Existing ML-based video quality enhancement mostly focuses on improving video resolution given the complete yet lower-resolution video frames. In practice, at the playout time, the receiver may not have a complete video frame to play either due to excessive delay or packet losses. Many measurement studies report a widely fluctuating delay in the Internet and wireless networks (\eg, \cite{constancy,5G-measurement1,5G-measurement2}). Meanwhile, packet losses are also very common. In a wireline network, losses happen due to queue drop during network congestion. 
In a wireless network, in addition to network congestion and large delay, losses also occur during low signal-to-noise ratio (SNR), collisions, and handoffs. For example, several measurement studies report some receivers experienced loss rates above 10\%~\cite{constancy,udp-loss}. \cite{5G-measurement2} reports during handoff the average latency increases by 2.26x and loss rates increase by 2.24x, which significantly degrades video streaming performance. \cite{5G-measurement1} reports the loss rates of 5G sessions are several-fold higher than 4G sessions. Moreover, adding a large buffer may prevent packet drop but lead to bufferbloat problem, which is prevalent in the Internet (\eg, \cite{bufferbloat1,bufferbloat2,bufferbloat3}), causes excessive delay, and harms video streaming performance.

Packet losses not only reduce the available data rate but also lead to a loss of a complete or partial video frame. While Forward Error Correction (FEC) and retransmissions have been widely used for loss recovery, their effectiveness is still quite limited. Retransmissions incur significant delay and may not be acceptable when round-trip time (RTT) is large.  FEC is expensive: as we show in Section~\ref{sec:motivation}, 35\% FEC is required in order to recover 5\% packet losses! Meanwhile, ML-based video prediction can potentially be used to conceal video errors or losses. However, their accuracy is limited since new content will appear in the next video frame, which makes it hard to predict. {\em Therefore, it is necessary to develop effective video recovery schemes.}

Moreover, various super-resolution (SR) algorithms have been developed to provide good video quality (\eg, \cite{chan2021basicvsr,yi2019progressive,sr3, sr4, sr5, sr6, sr7}). Yet despite significant research, the existing SR cannot support real-time execution on mobile devices. Interestingly, video streaming for mobile devices is becoming increasingly popular. \cite{mobile-video1} reports that 60\%+ U.S. digital video audiences watch videos using their smartphones. {\em This calls for the development of SR for mobile devices.} 

In addition, Adaptive video Bit Rate (ABR) has been widely used in the Internet to dynamically adjust the video streaming rate according to the current network conditions. Existing work adjusts the video rate based on what content has been received by the client. {\em As the clients increasingly adopt advanced video enhancement techniques, it is important to use the enhanced video to drive the design of ABR algorithms.}

\para{Our approach:} Inspired by the existing video enhancement and its limitation, in this paper we aim to advance the state of the art by developing (i) video recovery schemes, (ii) SR for mobile devices, and (iii) enhancement aware video recovery. A nice property of our design is that our approach works with existing video codecs and is easy to deploy. 

More specifically, for (i), we observe that simply predicting videos based on previously received data is error prone. If the sender could send some hint about the current videos in a reliable way, the receiver can use the hint to significantly improve the video prediction. The main challenge is to determine what hint to provide to optimize the recovered video. We develop a novel way to extract compact yet essential states from a video sequence and reliably transmit the video frame state (\eg, using TCP) for video recovery. Inspired by recent quantized image coding techniques~\cite{AdityaRamesh2021ZeroShotTG,WilsonYan2021VideoGPTVG}, we exploit the temporal locality in the video and employ a binary  point code to encode a frame into a very low-resolution binary point code.  
We find that the learned binary point code encodes both inter-frame movement information and contour information. Our experiments demonstrate that with the learned binary point code, the receiver can significantly improve the quality of video recovery. \yifan{In our system, we leverage TCP transmissions to deliver binary point code as auxiliary information to help  video recovery.} Our video recovery has the following distinct advantages: (i) Our binary point code is highly compact: within 1 KB, and yet significantly improves the video quality, (ii) it supports real-time extraction on the server and real-time video recovery on a mobile device, and (iii) it handles both partial and complete video frame recovery.

For (ii), we also advance the state-of-the-art in super-resolution by developing a novel video super-resolution algorithm at multiple resolutions. To address the limited computing power and resources on mobile devices, we use a single neural network model for all resolutions and leverage shared parameters to reduce memory overhead. At the bottom of the network, distinct structures have been designed to accommodate different resolutions (240p, 360p, 480p to 1080p) and provide the ABR algorithm with the flexibility to choose different rates. Different from the existing algorithms, it can support (i) different input video resolutions, and (ii) real-time execution on mobile devices (\eg, iPhone 12). 


For (iii), in addition to advancing receiver-side video enhancements, we further develop receiver-aware video bit-rate adaptation. Existing video rate adaptation selects the transmission video rate to maximize the video quality of experience (QoE), which consists of three major factors: video quality determined by the video data rate, change in the video quality, and rebuffering time. The video quality is determined based on the data transmitted to the receiver. Now that the receiver uses various enhancement techniques to improve the video quality, its actual video quality is likely to be much higher than the video directly received from the sender. Therefore a more effective approach is to use the enhanced video quality to drive the video bit rate decision. We develop an approach to efficiently estimate the impact of video recovery and SR on the video quality and rebuffering time, and use the estimation to facilitate bit rate selection. 

To understand the benefit of our video enhancement approaches and the enhancement-aware ABR algorithm, \zhaoyuan{we conduct our evaluation using QUIC~\cite{langley2017quic} under a variety of network conditions, including WiFi, 3G, 4G, and 5G networks on iPhone 12.}
\kj{Since iPhone 12 and higher versions account for more than 94\% of the total U.S. iPhone purchases in Q1 2023, reported in~\cite{CIRP}, the performance of most users' smartphones should exceed or be on par with that of the iPhone 12.}

Our major contributions can be summarized as follows:
\begin{sitemize}
    \item Our video recovery approach efficiently extracts binary point code from each video frame to better support video reconstruction at the receiver upon partial or complete video frame losses or excessive delay. 
    \item Our super-resolution algorithm further enhances the resolution of the video frames in real-time on mobile devices.
    \item Our video bit rate adaptation harnesses the full benefit of video recovery and super-resolution approaches by using the video QoE after enhancement for rate adaptation.
    \item Our evaluation in diverse types of networks shows that our approach enables real-time enhancement and improves the video QoE by 23.7\% - 51\%, 32.2\% - 68\%, 37.1\% - 82\%, and 29\% - 72\% in 3G, 4G, 5G, and WiFi networks, respectively. 
\end{sitemize}

This paper does not involve human subjects and has no ethical concerns. We plan to release our code and traces to the public. 
\section{Related Work}
\label{sec:related}

We classify existing work into the following three areas: (i) video recovery, (ii) video super-resolution, and (iii) ABR algorithms. 

\para{Video recovery: } Video recovery work can be categorized into two areas: FEC-based and ML-based approaches. FEC algorithms are frequently employed in video streaming systems (\eg, DASH, Apple HLS) to enhance video quality over unreliable networks. These algorithms add redundant information to the original video data, enabling the receiver to recover lost packets. Popular FEC algorithms include Reed-Solomon (RS) code~\cite{reed1960polynomial} for correcting burst errors, Low-Density Parity-Check (LDPC) code for efficient XOR-based error recovery, and Convolutional code for correcting errors spread across multiple bits. The choice of FEC algorithm depends on the specific requirements of the streaming system and the expected error types in the network.

Recently, many ML based algorithms have been proposed for video prediction, such as convolutional networks (\eg, \cite{cnn1,cnn2,cnn3,cnn4}), recurrent networks (\eg, \cite{rnn1,rnn2}), and generative networks (GANs). Among them, GANs are particularly effective for generation tasks. However, GANs are typically designed to generate realistic images, which may not be necessarily similar to the next video frame.  Several other works utilize neural networks to realize video error concealment such that corrupted frames can be recovered from previous frames. \cite{sankisa2018video} predicts optical flow from generated flows of past frames to reconstruct the degraded portion of the frame. \cite{sankisa2020temporal} designs a capsule network framework to extract the decoded temporal dependencies, which are further combined with the past frames and passed through a reconstruction network to perform motion-compensated error concealment.

\para{Video super-resolution: } There has been extensive work on designing Video super-resolution (SR) techniques. They leverage the previous several video frames to enhance the resolution of the current frame. BasicVSR~\cite{chan2021basicvsr} employs bidirectional RCNN where the features are propagated forward and backward independently. Moreover, PFNL~\cite{yi2019progressive} leverages the non-local sub-network to compute the correlations between all possible pixels within and across frames. In the motion estimation and motion compensation (MEMC) methods~\cite{sr3, sr4, sr5, sr6, sr7}, most of them adopt deep learning techniques to estimate the optical flow such that the motion information can be used to boost the quality of video super-resolution. NAS~\cite{yeo2018neural} is an adaptive streaming system that utilize scalable content-aware super-resolution DNNs. However, NAS uses desktop-class GPUs as clients and can only support on-demand video streaming. NEMO~\cite{yeo2020nemo} extends this idea to mobile devices, achieving real-time video enhancement. However, NEMO requires offline preparation for anchor point selection and separate model training, limiting its support to on-demand video streaming only. \zhaoyuan{PreSR~\cite{zhou2022presr} implements selective prefetching of video chunks for SR to enhance QoE, but it lacks real-time performance on mobile devices and involves significant optimization costs. DeepStream~\cite{amirpour2022deepstream} optimizes bitrate ladders and utilizes a lightweight SR model, but it still requires high-performance GPUs at the client side.} In 360-degree video streaming systems, ~\cite{chen2020sr360, dasari2020streaming} use SR techniques but do not support  general video streaming. \zhaoyuan{Dejavu~\cite{hu2019dejavu} leverages visual content similarity across video conference sessions to learn a mapping for real-time content enhancement. However, it necessitates offline learning using the sender's historical sessions, making it less effective for general video streaming where content changes are more substantial.}



\para{ABR algorithms:} In HTTP-based streaming (e.g. DASH, Apple HLS), videos are split into fixed-length chunks. Each chuck typically lasts 1-10 seconds each and is encoded at multiple bitrates. A client then uses an ABR to select a bitrate level for each chunk to take into account the amount of data already buffered locally and predict how long it takes to download a chunk at a given bitrate. A variety of ABR algorithms have been developed~\cite{akhtar2018oboe, huang2014buffer, jiang2012improving, mao2017neural, spiteri2020bola, yin2015control}. 
Among these approaches, MPC~\cite{yin2015control} and Pensieve~\cite{mao2017neural} demonstrate that directly optimizing for the desired QoE objective delivers better outcomes than heuristics-based approaches. In particular, Pensieve uses deep reinforcement learning and learns through “observations” how past decisions and the current state impact the video quality. Although these algorithms successfully cope with bandwidth variations, they do not consider the effect of client-side video enhancement.

NAS~\cite{yeo2018neural} and NEMO~\cite{yeo2020nemo} consider the impact of enhancement to some extent, but they do not provide detailed algorithms. 
Moreover, streaming DNN models introduces additional latency and is not able to provide an improvement for all video chunks. In comparison, our approach supports general videos and does not need offline DNN training for an individual video. Moreover, its ABR algorithm is aware of the impact of both video recovery and SR. 

\para{Summary:} Our work advances the state of the art in the following aspects: (i) we develop a novel video recovery scheme that extracts and leverages a compact state about the current video frame; compared with existing prediction schemes, which only use the previous video frames, our recovery can achieve higher accuracy with little overhead; (ii) we develop a new video super resolution scheme that achieves high video quality in real-time on mobile devices, whereas the existing SR works are too heavy weight for mobile devices; and (iii) we adapt video bit rate that explicitly optimizes the video QoE after applying video recovery and super-resolution; and (iv) we demonstrate the benefit of our approach under diverse network conditions, including WiFi, 3G, 4G, and 5G network conditions. 
\section{Motivation}
\label{sec:motivation}

In this section, we conduct controlled experiments to shed light on the limitations of existing FEC, super resolution, and video ABR algorithms. 

\begin{figure*}
\centering
\begin{minipage}[h!]{1\columnwidth}
\begin{minipage}{0.48\columnwidth}
    \begin{figure}[H]
    \centering
    \includegraphics[width=0.9\columnwidth]{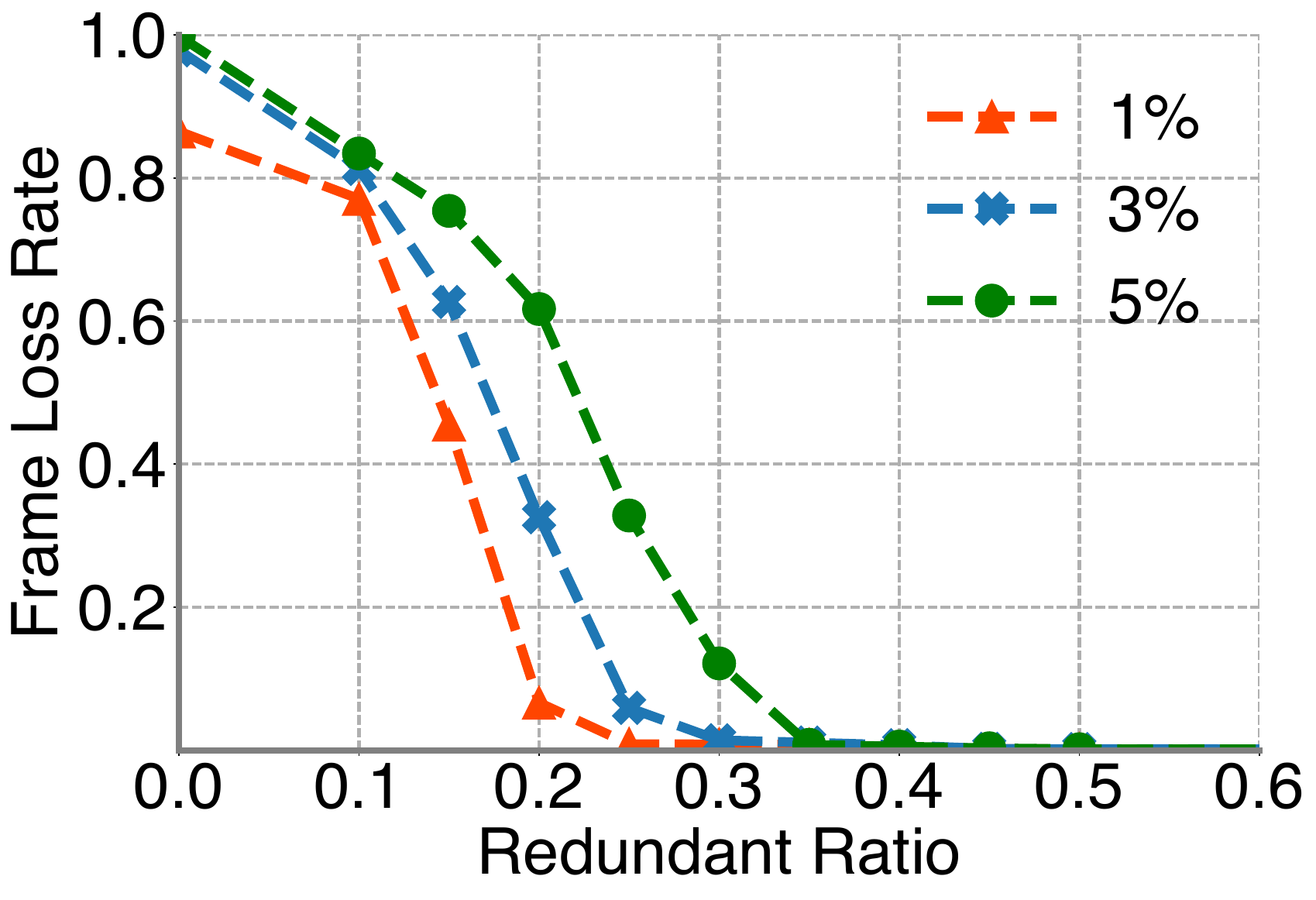}
    \caption{Frame loss rates under different packet loss rates and FEC redundancy ratios.}
    \label{fig:frame_loss_redundant}
    \end{figure}
\end{minipage}
\hspace{5pt}
\begin{minipage}{0.48\columnwidth}
    \begin{figure}[H]
    \centering
    \includegraphics[width=0.9\columnwidth]{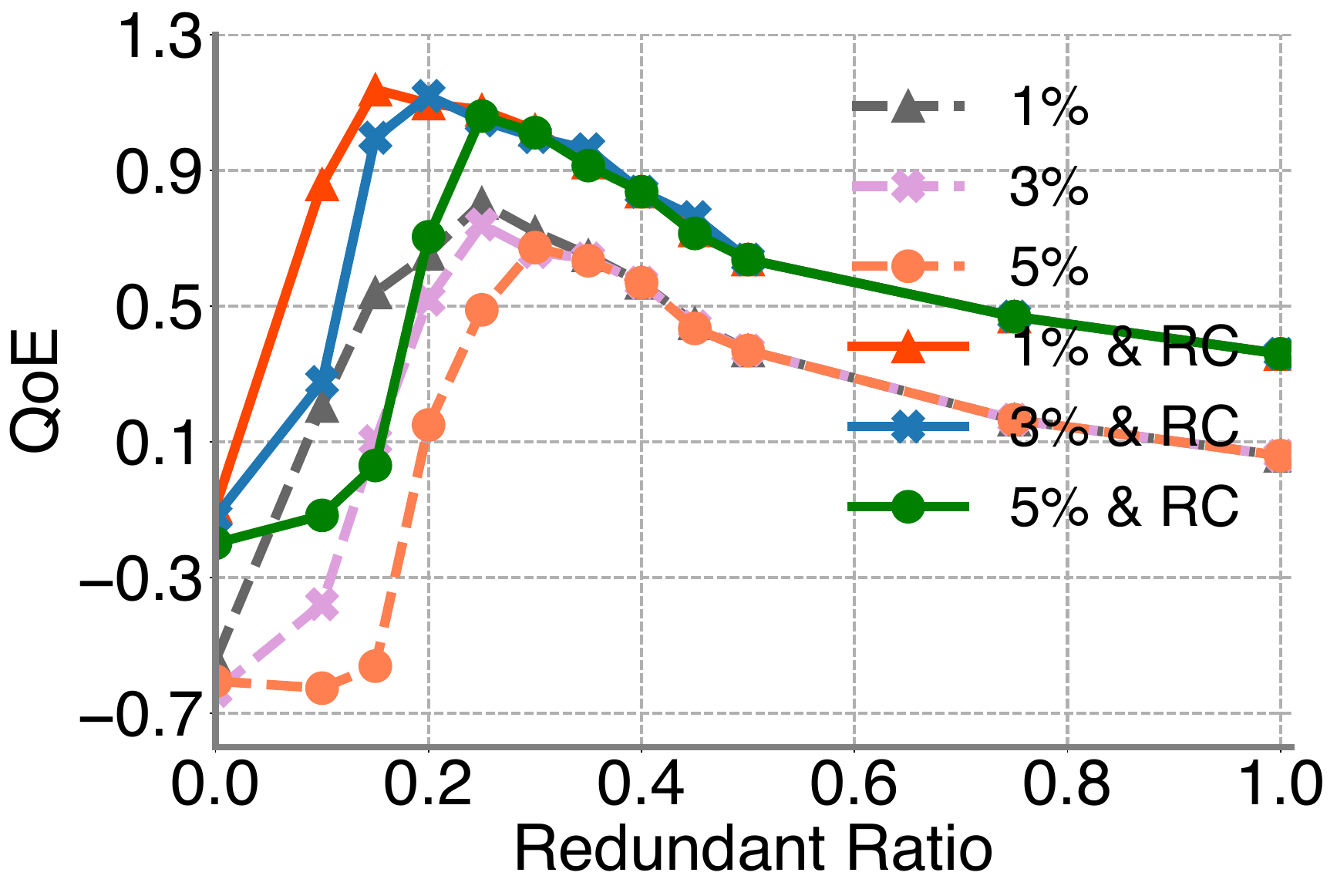}
    \caption{QoE with different FEC redundant ratios and with or without video recovery (RC).}
    \label{fig:qoe_redundant}
\end{figure}
\end{minipage}
\end{minipage}
\end{figure*}

\para{Limitations of FEC:} FEC has been widely used for video frame recovery. For example, WebRTC is a popular technology that supports steaming videos as well as data and voice on both browsers and native clients. It uses a hybrid of Negative Acknowledgement (NACK) and FEC for error recovery during video streaming. When the roundtrip time (RTT) is low, NACK is used; otherwise FEC is used to avoid excessive delay. We evaluate the benefit of FEC. \zhaoyuan{As shown in Figure~\ref{fig:frame_loss_redundant}, 1\% packet losses require 25\% FEC in order to achieve close to 0 video frame losses. The corresponding numbers for 3\% and 5\% packet losses are 30\% and 35\% FEC. These numbers show that FEC is very expensive. Similar observations are reported in \cite{RFEC}.}

Figure~\ref{fig:qoe_redundant} further plots the video QoE under different packet loss rates. \zhaoyuan{Under 5\% packet loss rate, the QoE even first decreases after adding FEC until sufficient FEC is added to recover the packet losses.} In general, FEC redundancy needs to be around 5x higher than the packet loss rates in order to support a successful recovery. This is rather expensive. These results motivate our work on developing ML-based video recovery. We also compare our video recovery under different amounts of FEC. We observe a similar trend as we increase the redundancy level. Meanwhile, our recovery achieves a higher QoE and reduces the amount of desired FEC. 



\begin{table}[]
\centering
\resizebox{0.55\columnwidth}{!}{%
\begin{tabular}{@{}ccccc@{}}
\toprule
\multirow{2}{*}{} & \multicolumn{4}{c}{Methods}       \\
                  & RLSP~\cite{DarioFuoli2019EfficientVS}   & BasicVSR~\cite{KelvinCKChan2021BasicVSRTS} & CKBG~\cite{JunXiao2022OnlineVS}  & ours  \\ \midrule
FLOPS(G)          & 132.94 & 71.33    & 17.8  & 10.8  \\
params(K)         & 1154   & 1887     & 1750  & 1619  \\
latency(ms)       & 5000   & 3500     & 1000  & 22    \\
PSNR              & 28.5   & 29.8     & 29.7  & 27.1  \\
SSIM              & 0.814  & 0.853    & 0.851 & 0.801 \\ \bottomrule
\end{tabular}%
}
\vspace{10pt}
\caption{Compare super-resolution with other methods. FLOPS and latency were validated on the REDS4~\cite{Nah_2019_CVPR_Workshops_REDS} dataset using 180*320 resolution as input for a 4x up-scale super resolution, produced on an iPhone 12.}
\label{tab:existing-sr}
\end{table}

\para{Limitations of existing super-resolution algorithms:} Table~\ref{tab:existing-sr} shows the performance and running time of existing SR algorithms on iPhone 12. We train on the REDS~\cite{Nah_2019_CVPR_Workshops_REDS} dataset and evaluate Peak Signal to Noise Ratio (PSNR)~\cite{PSNR} and Structure Similiarity (SSIM)~\cite{SSIM} on REDS4 for fair comparison. Both PSNR and SSIM are widely used video quality metrics, and their higher values indicate better quality. As we can see, the existing approaches are too slow for mobile devices. These results indicate the need for developing ML recovery and SR for mobile devices. Because we use smaller feature maps and more efficient feature utilization into optical flow network, as well as wrapping optimization for mobile devices, our FLOPS are smaller and can run in real time on the iPhone device. 

\section{Video Recovery}
\label{sec:recovery}

In this section, we present a deep neural network-based video recovery model for the client to recover the video whenever video frames get lost. This significantly enhances the resilience of video streaming under diverse network conditions. Note that there has been considerable work on video prediction, which predicts the next video frame based on the previous frames and can be applied to recover lost videos in our context. However, predicting solely based on the previous information has limited accuracy whenever the video frame has new content, which is common in reality. Our main idea is to send some hint about the next video frame along with the encoded video to facilitate video recovery. We design a light-weight mechanism to extract a compact hint for recovering the next video frame and show it can significantly enhance the quality of the recovered videos. Below we describe (i) how to extract a compact binary code at the video server and (ii) how to leverage the code for video reconstruction at the client. 




\para{Extracting binary point code: }
In order to extract meaningful information from a video frame to help the video recovery, our encoder borrows the concept of edge detection to preserve the contours of objects. We adopt PidiNET~\cite{pidinet} trained on BSDS500~\cite{BSDS500} as an edge encoder due to its stable performance and small inference overhead. The result of edge detection is usually a value between 0 and 1, so we binarize it here to form a binary point code.
The resolution of this binary point code can be very low. We find that even a $64\times128$ code (only 1 KB) can significantly improve the quality of the video recovery. We use TCP to reliably transmit it at a low cost (within one RTT). 

We train the encoder end-to-end with the decoder for better reconstruction. We use a binarization layer in Movement Pruning~\cite{sanh2020movement}. To ensure the encoder-decoder structure is differentiable, we skip the gradient from binarization, which directly pass the gradient before the quantization layer to the upper layer. 

Figure~\ref{fig:vis_recovery} illustrates the learned binary point code. It captures the motion and contour information of the current video frame. Despite its small size, it significantly improves the prediction accuracy. 




\para{Leveraging binary point code:} The client recovers the current frame based on the current binary point code, previous video frames, and optionally a portion of the current frame that is received correctly. 

\begin{figure*}[h!]
  \centering
  \includegraphics[width=\textwidth]{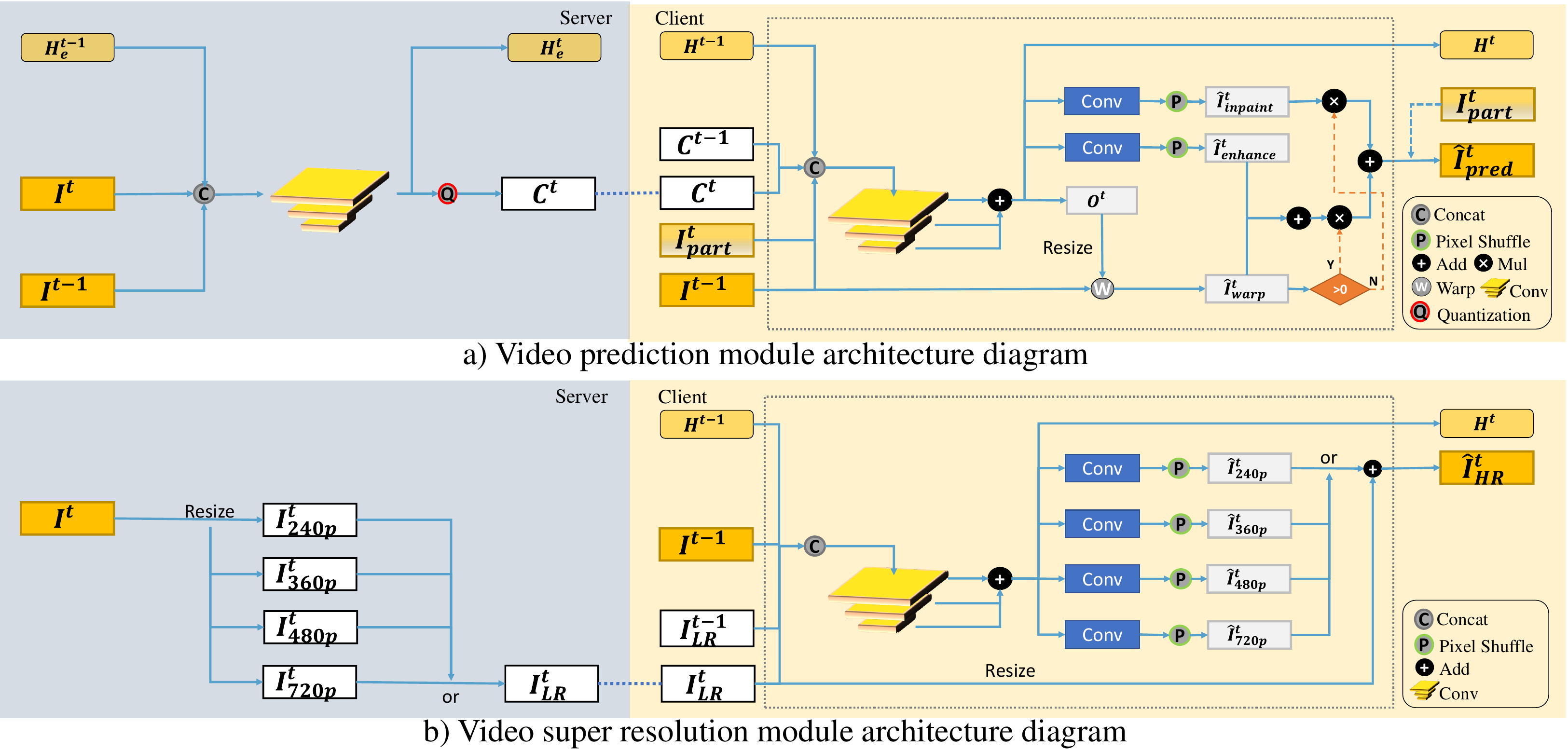}
  \caption{Model architecture. The blue portion shows the server side operation and yellow portion shows the client side. $I^t$ and $I^{t-1}$ are the video frames of the current time step $t$ and the previous $t-1$. $H_e$ is the history state maintained by the server-side encoder and $H$ is the history state maintained by the decoder. $C$ is the binary point code generated by the server, which is transmitted to the client to help with video recovery. $I^t_{part}$ is the partially decoded image when the transmission fails.  $O$ is the current optical flow. $I^t_{240}$, $I^t_{360}$, $I^t_{480}$ and $I^t_{720}$ are different resolutions of video frames downsampled from $I^t$. $I^{t}_{LR}$ is the input frame of the super-resolution network received from server, which is one of the four resolutions mentioned before, and will choose its corresponding upsampling structure to generate high-resolution prediction $\widehat{I}^t_{HR}$.}
  \label{fig:arch}
\end{figure*}





Its video recovery consists of the following steps: 
(i) estimating the optical flow between two consecutive binary point codes, (ii) upsampling the optical flow to the resolution of the video frame and warping the previous video frame to the current frame using the optical flow, (iii) further enhancing the content that is warped from the previous frame, and also using inpainting to generate new content that does not appear in the previous frames.

For (i), we use SpyNet~\cite{ranjan2017optical}, an efficient optical flow network, to derive the delta between the two binary codes. Different from SpyNet, we fine-tune the optical flow network end-to-end in order to warp the previous frame to match the current frame. 

For (ii), we find that warping on iPhone is very slow. We improve its speed by changing the wrapping resolution to 270p, which reduces the warping time to within 5 ms. 

For (iii), we improve the content in the warped regions based on the difference to the ground truth. Meanwhile, we observe that warp-based video frame prediction predicts the movement of the content that already exists in the previous video frames. We further use the binary point code from the current video frame to generate new content by upsampling it to the video frame size and using an inpainting module to generate the content in the region that is left empty after warping. It then concatenates the two prediction results to produce the final video frame. We use the Charbonnier loss~\cite{chan2021basicvsr}, a widely used loss function for generation tasks, as the optimization metric during training. 

To support partial error concealment, we introduce  $I_{part}$ as an additional input. $I_{part}$ is a partially decoded video frame due to packet loss, and the part they receive is a valuable input for recovering the missing part. We feed it to the recovery model to utilize this partial information, and partial content is also used to override the predicted $\widehat{I}_{pred}$ in thecorresponding region. Refer to  Figure~\ref{fig:vis_conceal} for $I_{part}$ and the recovery results for the missing region.

\para{Recovery model implementation details:} Figure~\ref{fig:arch}(a) shows the overall framework of our video recovery. The server transmits the binary point code to help video prediction at the client.  To improve computational efficiency, all inputs to the optical flow network are downsampled to $64\times128$. Outside the optical flow network, we resize the output to $270\times480$ and feed it to subsequent convolution layers. We use PixelShuffle~\cite{shi2016real} to upsample by 4x to produce 1080p output. Because wrapping operation on a mobile device is too costly, we resize the 1080p $I^{t-1}$ frame to 270p resolution and then perform wrapping to generate $\widehat{I}_{wrap}^t$. But it is hard to avoid the information loss due to downsample, so this degradation needs to be compensated by the enhancement module. So we additionally feed the 270p $I^{t-1}$ into the enhance convolution layers to compensate for the loss. \yifan{$H_e$ and $H$ are two hidden layer states maintained at the server and client, respectively, following a structure similar to RNNs. $H_e$ is responsible for capturing the temporal information during the server-side encoding of the binary point code, while $H$ records the temporal information associated with the video recovery process based on the binary point code. During the training phase, at each time step, both $H_e$ and $H$ selectively propagate the most valuable features to the subsequent time step. We have selected a set of feature maps from the output of the optical flow network and applied two groups of convolution layers with non-shared parameters. This results in two intermediate predictions, $\widehat{I}_{inpaint}$ and $\widehat{I}_{enhance}$. The former focuses on filling the gaps in $\widehat{I}_{wrap}$ created by the wrapping process (where the newly emerged content cannot be matched by the optical flow), while the latter concentrates on further adjusting $\widehat{I}_{wrap}$ (as wrap is only based on moving pixels from historical content, $\widehat{I}_{enhance}$ provides more subtle change predictions).}

\para{Joint FEC and video recovery:} So far, we are focused on designing a video recovery model. In practice, one can use both FEC and video recovery to cope with network losses and excessive delay. As shown in Figure~\ref{fig:qoe_redundant}, the best QoE performance is achieved when we add an appropriate amount of FEC. To determine the right amount of FEC to add under our video recovery, we take the video training traces (described in Section~\ref{ssec:eval-method}) and play it under different network loss rates, where a loss means the packet is either lost or not received in time. For each network loss rate, we apply different levels of FEC and perform video decoding and recovery as described above. We derive the resulting QoE and select the FEC that yields the highest QoE. In this way, we offline build a lookup table that specifies the best FEC level for each loss rate. During online running, we predict the loss rate for the next video chuck and index to the table using the predicted loss rate to determine the appropriate FEC redundancy to use. A similar approach can be applied to support other recovery methods. 





\section{Super-Resolution Videos}
\label{sec:video_sr}

We develop a DNN for super-resolution videos in real-time. As shown in Section~\ref{sec:motivation}, existing super-resolution videos are either too slow or not accurate. Our super-resolution can achieve good accuracy on mobile devices (\eg, iPhone 12) at 30ms per frame. 


Figure~\ref{fig:arch}(b) shows our Super-Resolution process. The network structure of our SR model and Recovery model are similar. Both are based on an optical flow network to align features, and then use upsampling modules to generate higher-resolution frames. To support streaming at different bit rates, the server will resize the video into different resolutions and transmit the resolution requested by the client as the $\mathbf{I}^{t}_{LR}$. On the client side, the optical flow network computes the pixel shift between $\mathbf{I}^{t}_{LR}$ and $\mathbf{I}^{t-1}_{LR}$. To save memory, SR models with different up-scaling will share the same optical flow network. To support super-resolution prediction at different resolutions, we use independent convolution layers to learn different degradation patterns. The learning target of SR network is the gap between the bilinear upsampled $\mathbf{I}^{t}_{LR}$ and the ground truth $\mathbf{I}^{t}$. As with the recovery model, we use Charbonnier loss to optimize above target and all scales task are trained simultaneously.

Our SR model stands out due to its unique properties, such as real-time execution on mobile devices and support for multiple input resolutions. This is achieved through an efficient design that incorporates a shared optical flow network for different up-scaling factors and independent convolution layers tailored to specific degradation patterns. Furthermore, our model provides up-sampling for different resolutions without incurring additional computational costs by resizing the feature maps to predict degradation at various resolutions. This innovative approach allows our model to effectively accommodate devices with diverse computational capabilities, ensuring an optimized video streaming experience across a wide range of devices under various network conditions.


\section{Enhancement Aware ABR}

We develop an enhancement aware ABR algorithm. It is built on the ABR in Pensieve~\cite{mao2017neural}, but advances Pensieve in two respects: (i) it is enhancement aware ABR, which considers the impact of video recovery and super-resolution, and (ii) it incorporates the latest Reinforcement Learning (RL) algorithm -- Proximal Policy
Optimization (PPO)~\cite{PPO}. Below we focus on how to model the impact of video recovery and super-resolution on video QoE, and refer the readers to ~\cite{PPO} for detailed description about PPO.

\begin{figure}[h!]
  \centering
  \subfigure[Relationship between PSNR degradation and consecutive recovered frames.]{%
    \includegraphics[width=0.4\columnwidth]{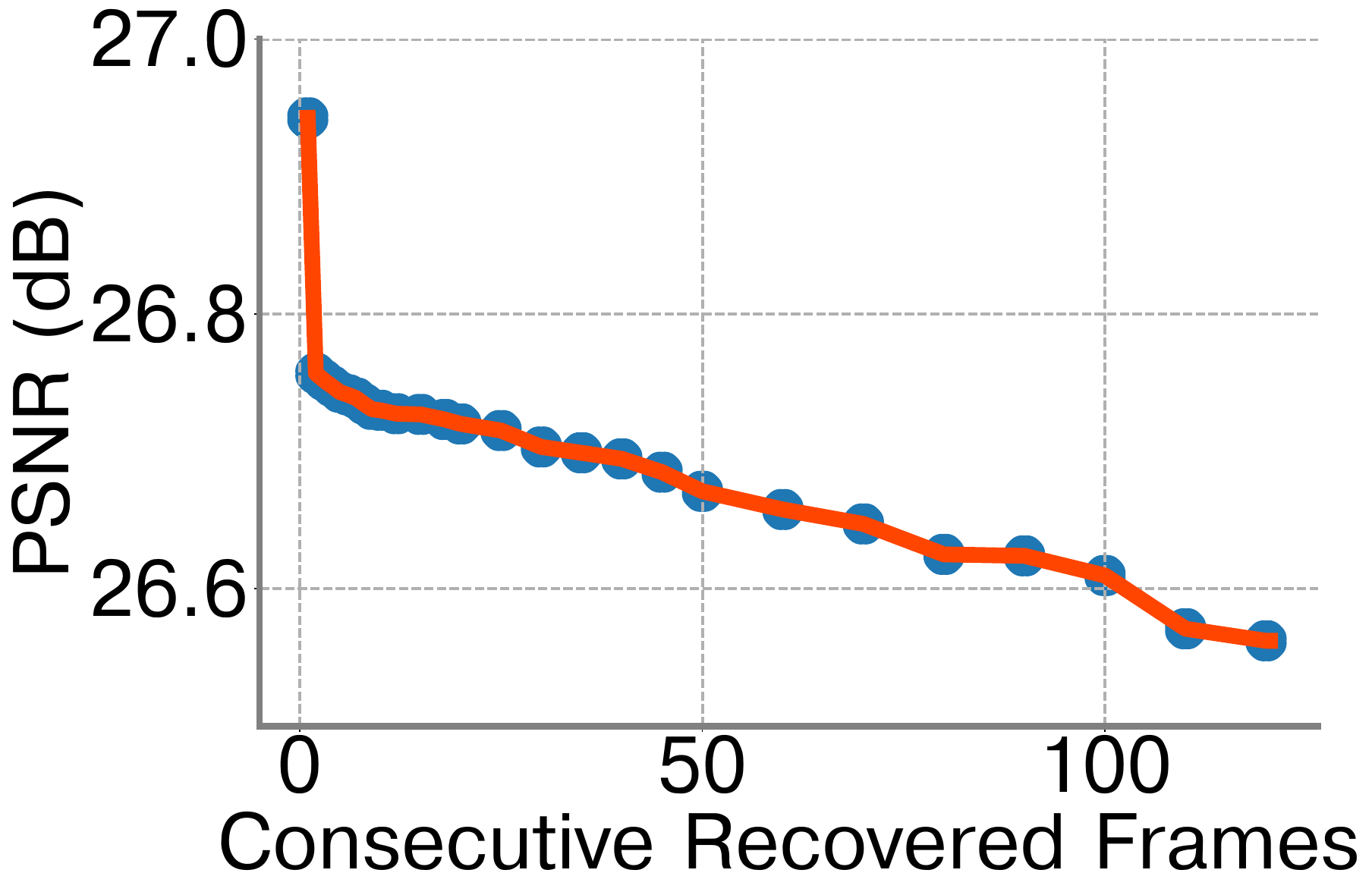}
    \vspace{-10pt}
    \label{fig:rcframes_to_psnr_map}
  }
  \hspace{5pt}
  \subfigure[Relationship between PSNR and Bitrate.]{%
    \includegraphics[width=0.4\columnwidth]{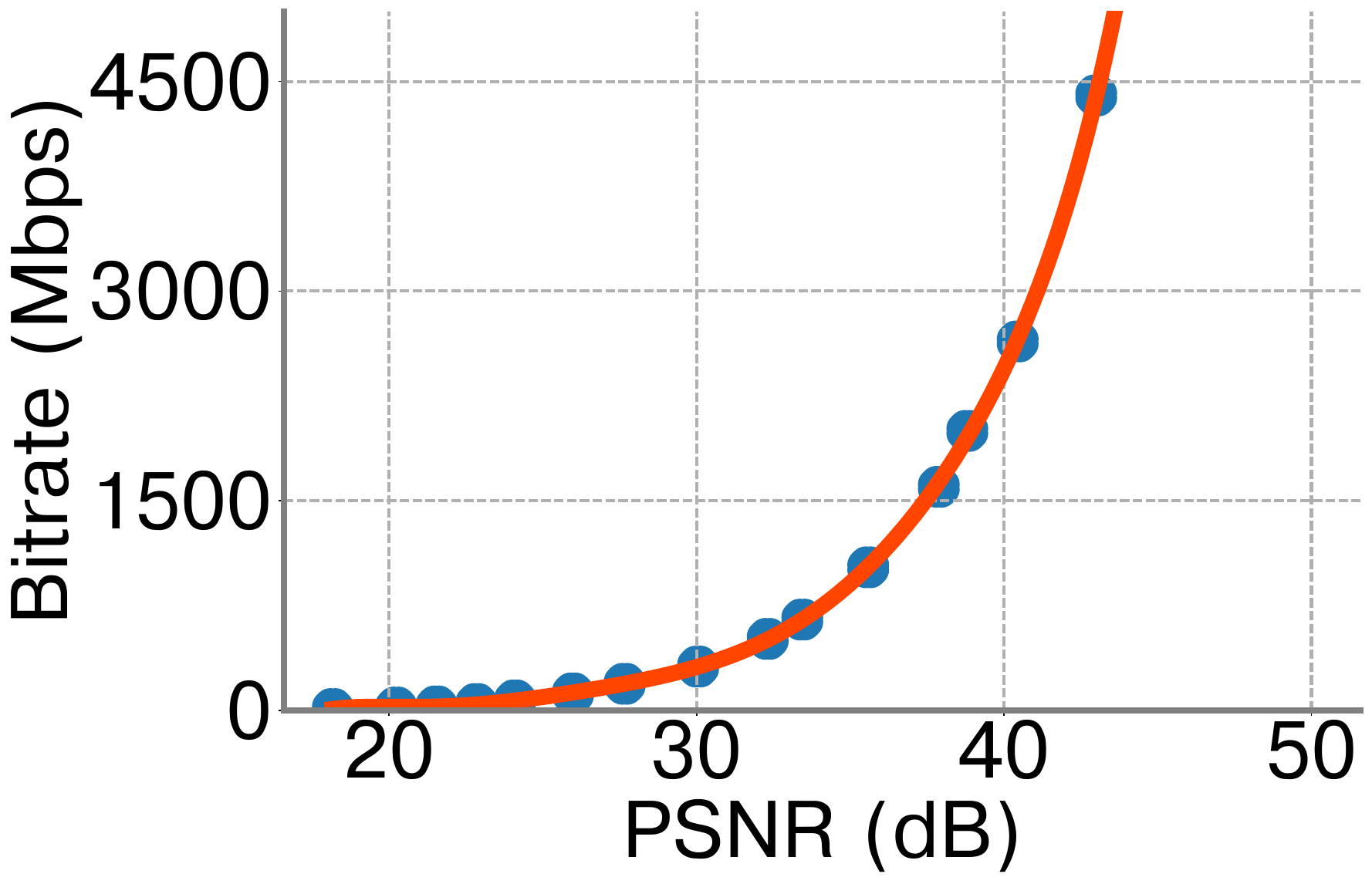}
    \vspace{-10pt}
    \label{fig:bitrate_to_psnr_map}
  }
  \caption{Mapping functions for the impact of video recovery and video super-resolution on video quality.}
  \label{fig:mapping}
\end{figure}

\para{Quality of Experience (QoE):} The widely used video Quality of Experience (QoE) is a function of bitrate utility, rebuffering time, and smoothness of selected bitrates, which is defined as follow:
\[\frac{\sum_{n=1}^{N} R_{n} - \mu \sum_{n=1}^{N} T_{n} - \sum_{n=1}^{N-1} |R_{n+1} - R_{n}|}{N} \]
where $N$ is the number of video chunks, $R_{n}$ and $T_{n}$ represent the video chunk n’s bitrate and the
rebuffering time resulting from its download, respectively, and $\mu$ is the
rebuffering penalty. Video recovery and super-resolution impact the QoE in two ways: (1) they improve the video quality and its corresponding bitrate and (2) their running time also affects the rebuffering time. Below we compute (i) how video recovery affects the video quality, (ii) how SR affects the video quality, (iii) how video recovery affects rebuffering time, and (iv) how SR affects rebuffering time. 

\para{Impact of video recovery on video quality:} Let us first consider how video recovery affects the video quality. The impact can be quantified based on how many video frames go through video recovery and how much video quality change after recovery. First, we estimate the number of video frames in the current video chuck that need recovery. We observe that recovery takes place when either (i) a video frame is lost (\eg, due to network congestion, low SNR, mobility) or (ii) a video frame arrives too late (\ie, the frame has not arrived by the time the video frame is scheduled to play). We can predict the loss rate (\eg, using Exponential Weighted Moving window Average (EWMA) or Holt Winters (HW)), and use the predicted loss rate to estimate (i). To estimate (ii), we compute the expected play time for the $i$-th video frame as $T_{play} = T_{prev} + i \Delta$, where $T_{prev}$ is the end time of the previous video chuck, $i$ is the video frame index, and $\Delta$ is the inter-frame time. We compute the expected arrival time for the $i$-th frame as $T_{arr}=T_{prev} + \sum_i S_i/tput_{curr}$, where $S_i$ is the total size of data for the $i$-th frame and $tput_{curr}$ is the predicted throughput for the current video chuck, which can be derived using EWMA or HW. Then for every video frame $i$ in the current chuck, we count the number of frames whose $T_{play}^i < T_{arr}^i$. 

Next we need to compute the video quality for the recovered frame. This depends on the selected video bit rate. For simplicity, we take videos from the top ten popular categories~\cite{medium-report} on YouTube as our training data; for each bit rate, we compute the average PSNR of these video frames after applying video recovery. We use this value as the estimate for the video quality. For the remaining videos that do not need recovery, we can estimate the video quality based on the selected bit rate (\eg, computing the average PSNR of our training video frames at that rate). 

\para{Impact of SR on video quality:} The impact of SR can also be quantified based on (i) the number of video frames that go through SR and (ii) the enhanced video quality after applying SR. (ii) can be estimated by computing PSNR of the frames in our training videos after applying SR at each bit rate. 

Below we compute (i). Since rebuffering time is more annoying than degraded video quality, we skip SR if SR can cause rebuffering. Therefore video frames can be grouped into the following three scenarios: 1) those that are not received in time for playout and require recovery, 2) those that are received in time and can be applied SR before the playout time, and 3) those that are received in time but cannot finish SR before the playout time. To derive 2), for every video frame $i$ in the current chuck, we count the number of frames whose $T_{play}^i > T_{arr}^i + T_{SR}$, where $T_{SR}$ is the processing time of SR.


\para{Impact of video recovery on rebuffering time:} To quantify the video recovery on the rebuffering time, we observe that only the frames that go through recovery has rebuffering time. \zhaoyuan{These frames can be recovered either when they are received successfully but late or corrupted due to network losses.} Therefore, their rebuffering time can be estimated as $min(\sum_i S_i/tput_{curr}-T_{play}^i, T_{RC})$, where $T_{RC}$ is the recovery time.

\para{Impact of SR on rebuffering time:} As mentioned earlier, to minimize rebuffering time, only the frames that can finish SR before their playout time will go through SR. Therefore, SR does not affect rebuffering time. 

\para{Enhancement-aware ABR:} Our final ABR algorithm computes the QoE of the current video chuck for each video bit rate and selects the rate that leads to the highest QoE.

\comment{
\subsection{Recovery Aware ABR}

\para{Video quality:} Let us first consider how video recovery affects the video quality. Our video recovery takes place when either (i) a video frame is lost (\eg, due to network congestion, low SNR, mobility) or (ii) a video frame arrives too late. 

While our video recovery handles both types of losses, our recovery-aware ABR algorithm only considers (ii) because the losses in (i) are typically low and it is challenging to predict exactly which video frame will be lost and need to go through video recovery. 

For the next video chuck, our goal is to estimate (1) how many video frames in the next video chuck needs recovery and (2) what is the video quality after recovery. For (1), we focus on (ii) and observe the $i$-th video frame in the chuck has an expected playout time $T_{playout}$, which can be calculated as $T_{playout} = T_{prev} + i \times \Delta$, where $T_{prev}$ is the end time of the previous video chuck, $i$ is the video frame index, and $\Delta$ is the inter-frame time. For each video frame $i$, we compute the buffer occupancy as $B_{curr}^i = max(0,B_{prev} - \sum_{k=1..i-1} D_i/ \Delta + i)$, where $B_{prev}$ is the buffer

\zhaoyuan{Zhaoyuan: We can consider the buffer occupancy as a buffer time to avoid the integer rounding problem, then the recovery-aware buffer time would be as follows.
\[B_{curr}^i = max(0, B_{prev} - \sum_{k=1}^{i} D_k + i\times \Delta)\]
$D_k$ would be determined by the transmission delay of the kth frame. The SR-aware buffer time would be as follows.
\[B_{curr}^i = max(0, B_{prev} - \sum_{k=1}^{i} D_k + i\times \Delta)\]
\[D_k = T_k + T_{sr}\]
Here $D_k$ would be the sum of the transmission delay and the SR inference time of the kth frame.
}

(i-1) + (i-1)\times \Delta + \sum_k tx_{k}$. For any video frames whose $B_{curr}^i$
goes below 0, they need video recovery.

For (2) 
need to compute the buffer occupancy at 

NAS defines an effective bitrate $R_{effective}$ as the bitrate that corresponds to the improved video quality (\eg, SSIM or PSNR). The mapping from video quality to video bitrate is a piece-wise linear interpolation. We follow the same idea but use a polynomial mapping function because it fits our data better. Our recovery can also provide a good effective bitrate $R_{effective}$, leading to a stable and high QoE. 

$R_{effective}$ depends on the recovered video quality, which degrades with an increasing number of recovered frames because we always use a predicted frame to further predict the next frame. In this case, the prediction error is accumulated, which degrades the recovery quality. To estimate the degradation, we first do DNN inference for our recovery model with different consecutive recovered frames and get the average PSNR among all the videos. Then we use interpolation to map the number of recovered frames to the average PSNR among these frames. The average PSNR can be further mapped to $R_{effective}$.

\para{Rebuffering time:} Next we examine how video recovery affects the rebuffering time. In on-demand video streaming, a video is divided into multiple video chucks, each lasting a few seconds. A video chuck can be played once it is successfully decoded. Rebuffering happens when we need to play out video but the video buffer is empty. In order to compute the rebuffering time, we need to first determine the expected playout time for each video frame and then check if the buffer is empty at the playout time. 

compute the rebuffering time after applying the video recovery. In general video streaming systems, as a client receives a video frame, the frame will be stored in a video buffer to be played later. The total delay of a video chunk is determined by the video transmission time. After the video chunk is fully received, we calculate the rebuffering time $T_{rebuf}$ and the buffer time $T_{buf}$ as follows. 
\[T_{rebuf} = max(delay - T_{buf}, 0)\]
\[T_{buf} = max(T_{buf} - delay, 0)\]

In our recovery-aware ABR, it is essential to find a recovery frame index and the corresponding total delay such that the QoE can be optimized. Our approach is described as follows. 

If we find the total transmission delay of the current video chunk larger than the current buffer time, then we perform Algorithm ~\ref{alg:get_recovery_delay} to get a recovery frame index and the corresponding delay to avoid the rebuffering penalty. We first acquire a list of transmission delays $Dl$ for all received video frames. Then, we iterate over all frames backward to find a video frame whose delay $T$ is less than or equal to the buffer time. $T$ is calculated based on the sum of the transmission delay until the frame is received and the recovery inference time of all remaining frames in the video chunk. Then, the recovery index $n$ and the corresponding delay $T$ would be returned. 

\begin{algorithm}[h!]
	\caption{Get recovery frame index of each video chunk}
	\label{alg:get_recovery_delay}
	\begin{algorithmic}[1]
    \State $N:$ The number of frames in the current video chunk
    \State $T_{buf}:$ The buffer time of the current video chunk
    \State $Dl:$ The list of transmission delays until each received frame
    \For{$frame\ f_{n},\ n = N,N-1\ldots,1$}
        \State Get the transmission delay $Dl[n]$ until frame $n$ is received; Get the recovery inference time $RC_{f_{n-1}}$
        \State $T$ = $Dl[n]$ + $RC_{f_{n-1}} \times (N - n)$
        \If{$T$ <= $T_{buf}$}
            \State \textbf{Return} $n$, $T$
        \EndIf
    \EndFor
	\end{algorithmic}
\end{algorithm}

\subsection{SR Aware ABR}

The goal of our SR-aware ABR is to maximize QoE by improving the effective bitrate $R_{effective}$ with our video super-resolution model while keeping the SR inference time low. After a client receives a video frame, the video super-resolution is performed. In the meantime, the client can receive the next video frame to be stored in the video buffer. In this case, a pipeline is formed such that the total delay of a video chunk is determined by the video transmission time and super-resolution delay together. The total delay is calculated by Algorithm~\ref{alg:get_sr_delay} and $D_{SR}$ would be the returned value.  

\begin{algorithm}
	\caption{Get delay of each video chunk}
	\label{alg:get_sr_delay}
	\begin{algorithmic}[1]
    \State $N:$ The number of frames in the current video chunk
    \State $D_{T}:$ Transmission delay of the current video chunk
    \State $D_{SR}:$ SR delay of the current video chunk
    \State Get the transmission time $T_{f1}$ of $frame\ 1$ from throughput and frame size
    \State $D_{T} = T_{f1}$
    \State $D_{SR} = T_{f1}$
    \For{$frame\ f_{n},\ n = 2,3,\ldots,N$}
        \State Get the transmission time $T_{f_{n}}$ from throughput and frame size; Get the SR inference time $SR_{f_{n-1}}$
        \State $D_{T}$ += $T_{f_{n}};\ D_{SR}$ += $SR_{f_{n-1}}$
        \If{$D_{T}$ > $D_{SR}$}
            \State $D_{SR}$ = $D_{T}$
        \EndIf
    \EndFor
    \State $D_{SR}$ += $SR_{f_{N}}$
	\end{algorithmic}
\end{algorithm}

For each frame of the current video chunk, we perform video super-resolution after the frame is received. At the same time, the next frame is being received. Following this pipeline, we receive the first frame before performing video super-resolution. Then, we get the transmission time of the current frame and the SR inference time of the last frame. If the accumulated transmission delay $D_{T}$ is greater than the accumulated SR delay $D_{SR}$, that means the low throughput takes up most of the overhead, thereby delaying the time of performing SR inference. If $D_{T}$ is less than $D_{SR}$, the transmission delay can be accumulated ahead of the SR delay. After all of the video frames are received, the total delay would be the maximal accumulated delay through the pipeline, namely, $D_{SR}$. 
}
\section{System Implementation}
\label{sec:system}

\begin{figure}[t!]
  \centering
  \vspace{-15pt}
  \includegraphics[width=0.95\columnwidth]{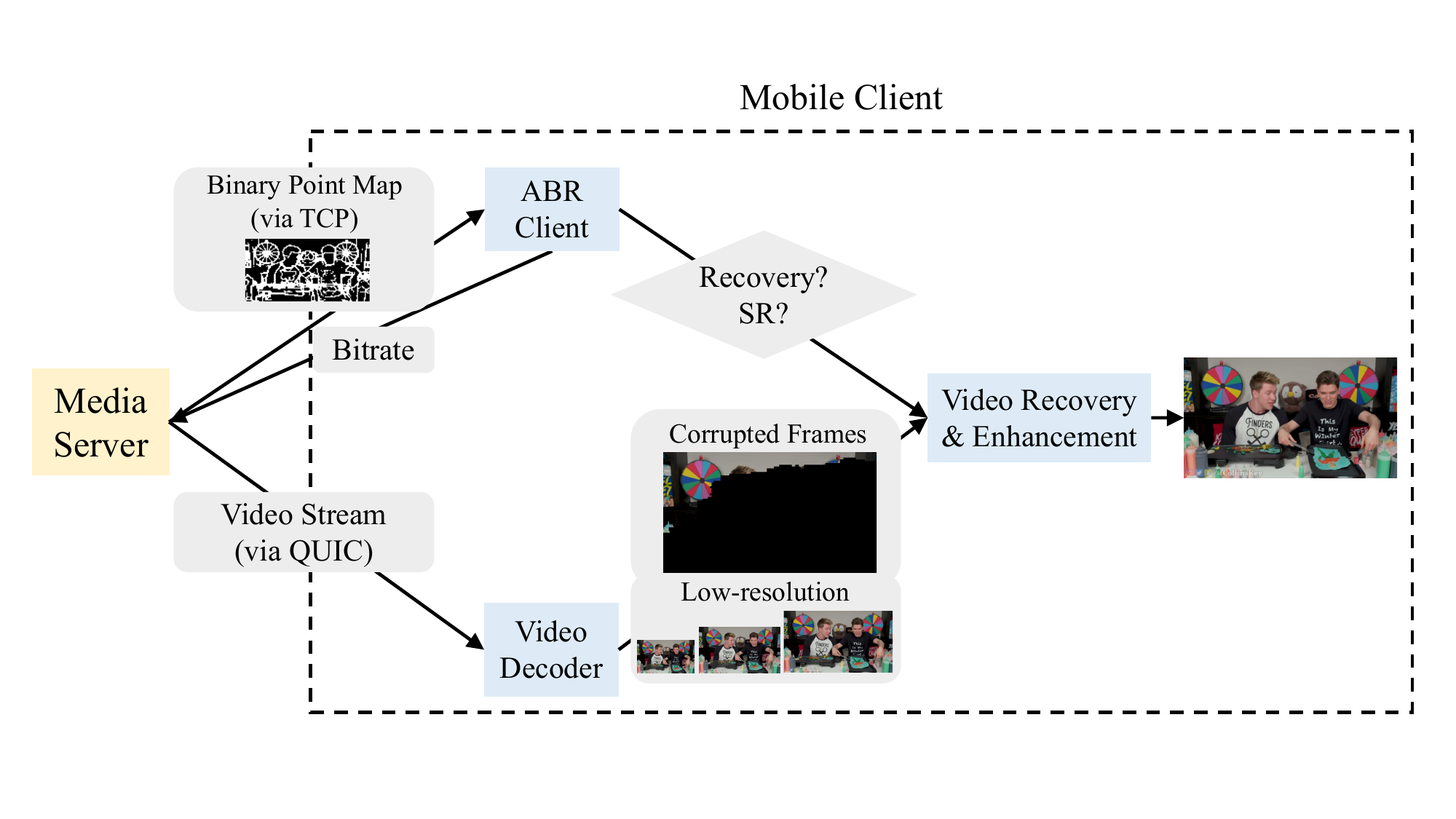}
  \vspace{-30pt}
  \caption{Overview of our system architecture.}
  \label{fig:sys_arch}
\end{figure}

Figure~\ref{fig:sys_arch} shows our system architecture.

\para{Server: } Our video server runs a standard video codec (\eg, VP9, H264 or H265) to encode a video as usually. Meanwhile, it extracts a binary point code and reliably transmits the code to the client via TCP. \zhaoyuan{Video contents are streamed using QUIC. QUIC~\cite{langley2017quic} has been a popular transport protocol for video streaming (\eg, YouTube, Instagram) as well as other applications, such as Gmail and Uber due to its more efficient connection setup and faster retransmission than TCP. Google reported that QUIC could reduce the latency for Google search by more than 2\% and the rebuffer time for YouTube by 9\% compared to TCP \cite{chrome-quic}. We find YouTube uses QUIC to stream recently uploaded and popular videos via Wireshark. QUIC has fast retransmission logic, but it still has 1.6\% packet loss in 5G networks.
} 

\para{Client: } Our client runs {\em our} enhancement-aware ABR algorithm to select the video bit rate and request the selected rate from the server. It runs a standard video decoder to decode the video frames as usual. After successful decoding, it applies {\em our} SR to enhance the video resolution and add the enhanced video frame to its buffer. Whenever a video frame needs to be played but its content is not completely available, it runs {\em our} video recovery to recover the video and feeds it to the player. \zhaoyuan{We use an iPhone 12 
as the mobile client in this paper.} 

\para{Model deployment: } Since we perform the video enhancement on the receiver side, we need to deploy our SR and recovery models to a mobile device. CoreML is commonly used because it optimizes on-device performance for iOS by leveraging the CPU, GPU, and Neural Engine. We observe that our model with the format of CoreML runs faster than ONNX, Pytorch Mobile, and TensorFlow Lite. However, the grid sample operation of warping runs slowly because it does not officially support GPU. To address this issue, we leverage Metal Performance Shaders (MPS) to create a custom grid sample layer running on GPU. MPS is a framework with handy Metal compute kernels and CoreML uses it for model inference on GPU. It also provides us with many APIs to create a custom layer so that the grid sample is implemented with a GPU acceleration. \zhaoyuan{In addition, we perform warping at a smaller scale of 270p instead of 1080p, thereby reducing the warping time from $29ms$ to $5ms$ on the iPhone 12. 
We use FP16 precision for both inputs and model weights without performance degradation to further reduce the inference time. The final total inference time is $22ms$.}
\section{Evaluation}
\label{sec:eval}

In this section, we first present our evaluation methodology and then describe performance results.

\subsection{Evaluation Methodology}
\label{ssec:eval-method}

\para{Network traces:} To understand the performance of our video enhancement approaches under diverse scenarios, we collect network traces over QUIC from WiFi, 3G, 4G, and 5G networks, shown in Table~\ref{tab:network_traces}. \kj{We use \textit{net-export}~\cite{net-export} in Chrome to collect the QUIC-related packets while watching Youtube videos. We especially identify packet loss in QUIC by capturing \textit{LOSS\_RETRANSMISSION} and \textit{PTO\_RETRANSMISSION} on the transmission type of the packet. Meanwhile, we measure the downlink throughput using iperf from an Azure server located in the central U.S. to a local client over the Internet. The 3G, 4G and 5G traces include static and walking scenarios. We also move the local client randomly to add mobility to the WiFi traces.}



\begin{figure*}[h!]
  \centering
  \includegraphics[width=1\columnwidth]{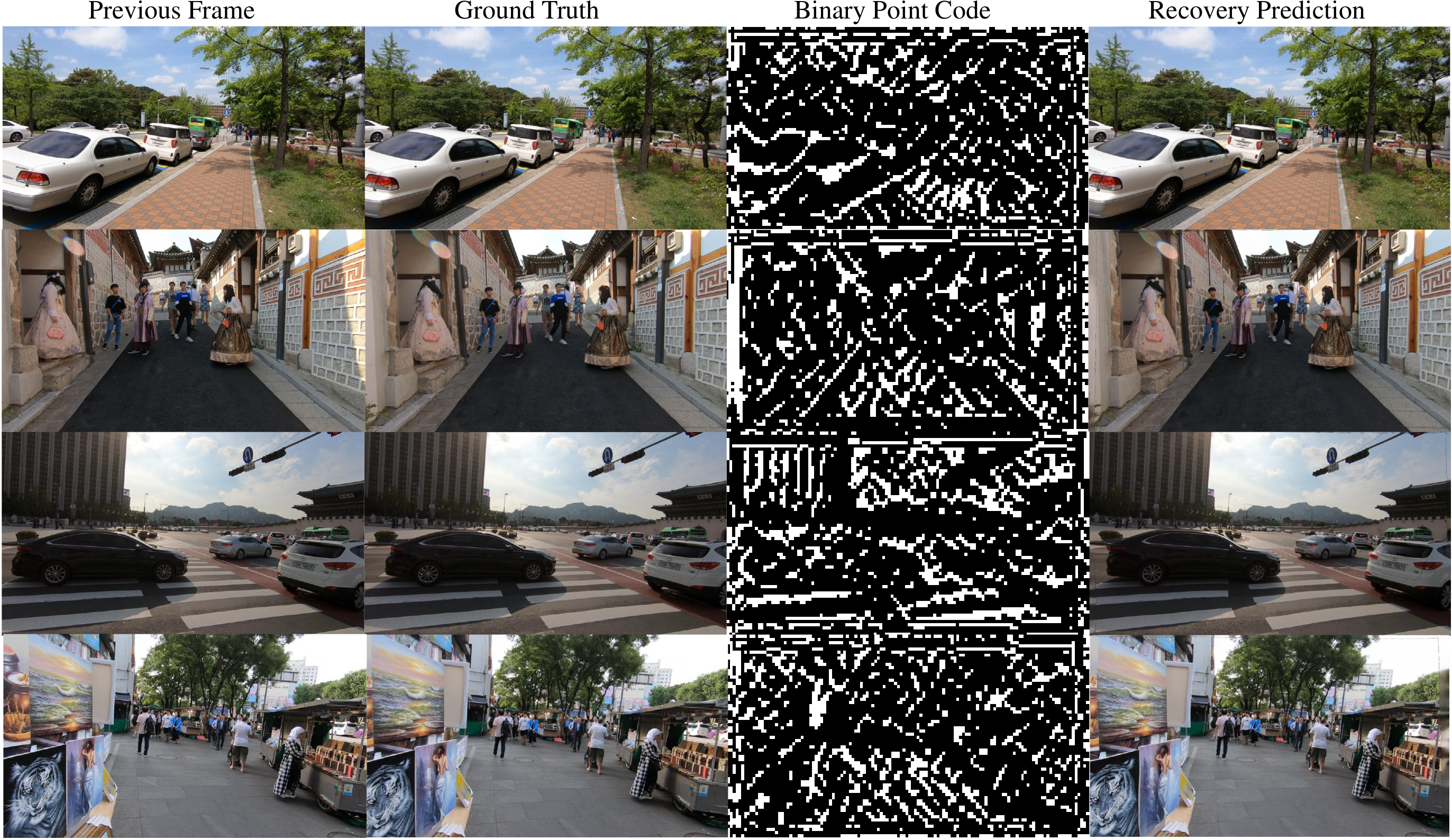}
  \caption{Visualization of video recovery results. Cases are sampled from REDS4~\cite{Nah_2019_CVPR_Workshops_REDS}.}
  \label{fig:vis_recovery}
\end{figure*}

\begin{table}
  \centering
  \resizebox{0.5\columnwidth}{!}{%
  \begin{tabular}{c|c|c|c|c}
    \toprule
    & 3G & 4G & 5G & WiFi \\
    \midrule
    Amount & 45 & 62 & 53 & 68 \\
    Avg. Duration (s) & 322 & 317 & 302 & 309 \\
    Avg. Throughput (Mbps) & 7.5 & 21.6 & 36.4 & 82.3 \\
    Avg. Packet loss rate (\%) & 0.9 & 1.3 & 1.6 & 0.5 \\
    \bottomrule
  \end{tabular}
  }
  \vspace{10pt}
  \caption{Network traces}
  \label{tab:network_traces}
\end{table}

\para{Video datasets:} We use the video dataset from NEMO for the evaluation. We choose videos from the top ten popular categories~\cite{medium-report} on YouTube: 'Product review', 'How-to', 'Vlogs', 'Game play', 'Skit', 'Haul', 'Challenges', 'Favorite', 'Education', and 'Unboxing'. From each category, we select five videos from distinct creators which support 4K at 30fps and are at least 5 minutes long. Then, four of them are distributed to the training dataset and the other belongs to the testing dataset. For adaptive streaming, we transcode them into multiple bitrate versions using the VP9 codec as per Wowza’s recommendation~\cite{wowza-recommendation}: \{512, 1024, 1600, 2640, 4400\} kbps at \{240, 360, 480, 720, 1080\}p resolutions. The GOP size is 120 (4 sec). 

\para{Performance metrics:} We use raw 1080p videos as a reference for measuring PSNR. We quantify the quality of recovered and super-resolved video frames using two widely used video quality metrics: SSIM and PSNR. Higher SSIM and PSNR values indicate better video quality. We quantify the performance of our system using QoE. A higher QoE indicates better video streaming for users.

\subsection{DNN Performance} 

First, we evaluate the DNN performance in terms of video quality.

\begin{figure*}
\centering
\begin{minipage}[ht]{1\columnwidth}
\begin{minipage}{0.5\columnwidth}
    \begin{figure}[H]
    \centering
    \includegraphics[width=1\columnwidth]{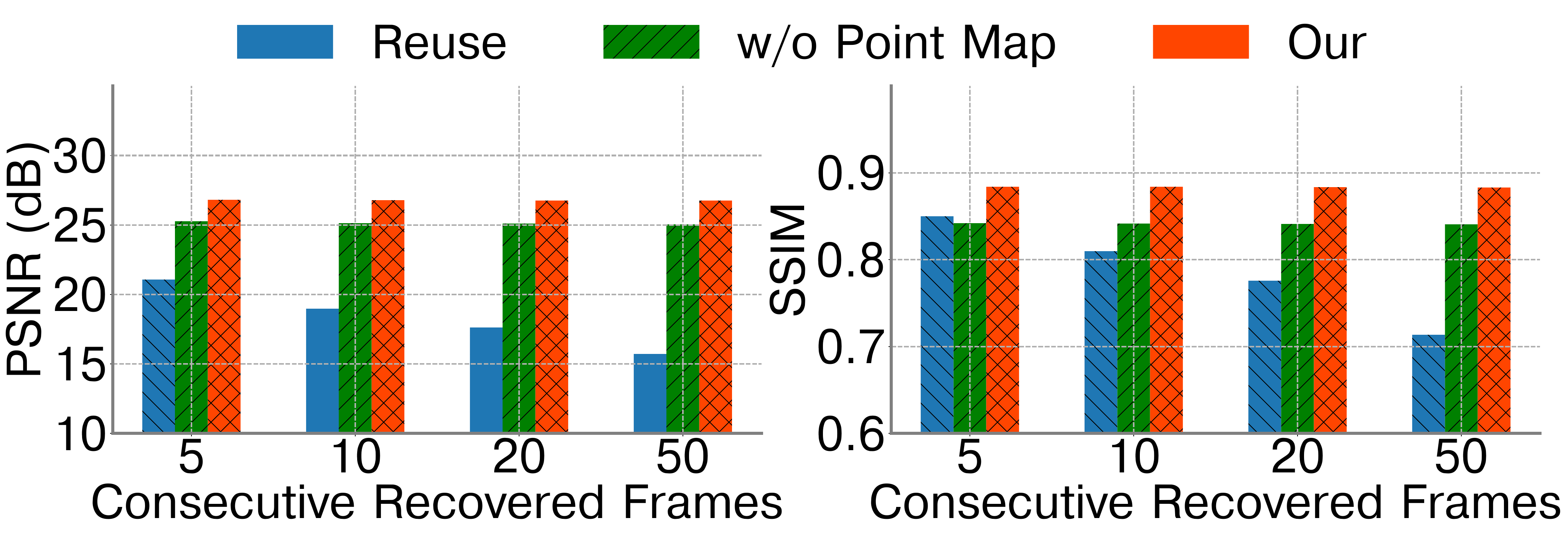}
    \vspace{-10pt}
    \caption{Performance results of video prediction.}
    \label{fig:rc_dnn_quality}
    \end{figure}
\end{minipage}
\begin{minipage}{0.5\columnwidth}
    \begin{figure}[H]
    \centering
    \includegraphics[width=1\columnwidth]{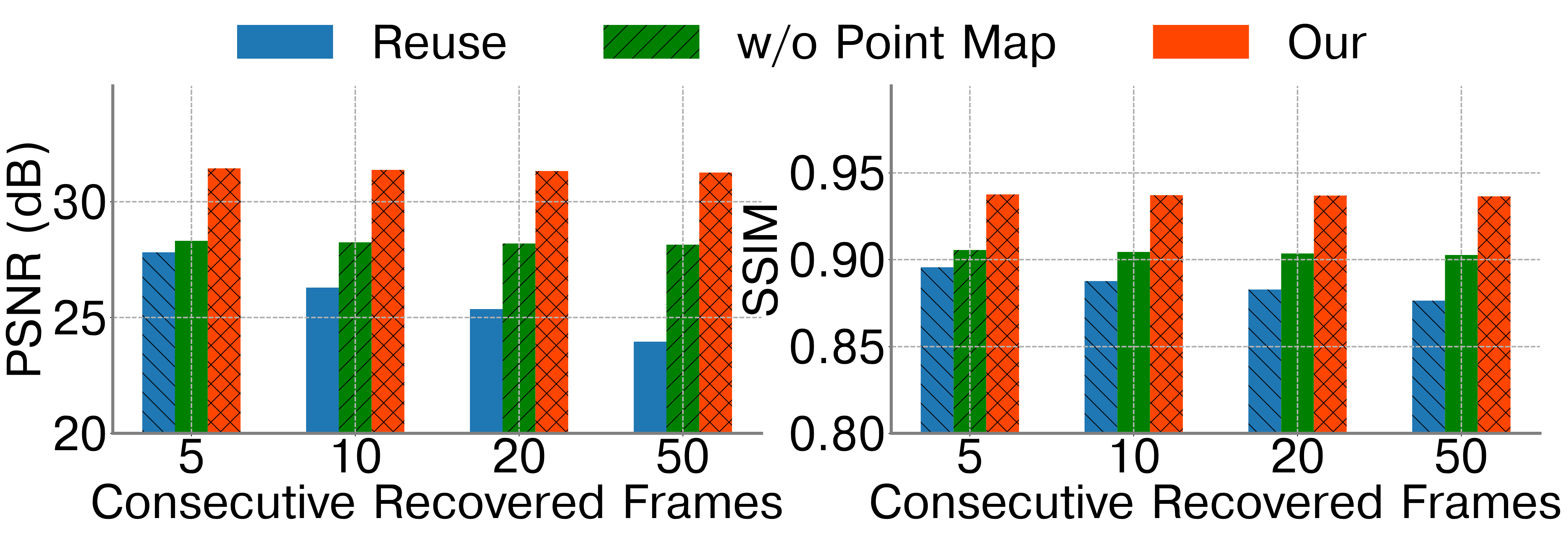}
    \vspace{-10pt}
    \caption{Performance results of video partial recovery.}
    \label{fig:rc_cl_dnn_quality}
\end{figure}
\end{minipage}
\end{minipage}
\end{figure*}

\para{DNN performance of video recovery: } Figure~\ref{fig:rc_dnn_quality} compares the video quality of simply reusing the previous video frame, predicting the video frame without the binary point code, and predicting using our binary point code. We use these schemes to predict the next 5, 10, 20, and 50 frames and calculate the average video quality. As we can see, video recovery without the binary point code yields 4-9dB PSNR improvement and 0.03-0.13 SSIM improvement over simple frame reuse; and the binary point code further increases PSNR by 6-12dB and increases SSIM by 0.04-0.17. The result shows the effectiveness of our binary point code. As we increase the number of future frames to predict, the prediction quality  using our recovery model degrades gracefully. 

Figure~\ref{fig:vis_recovery} further illustrates the visualization of our video recovery results. As we can see, our recovery model can learn the motion movement between two consecutive frames and the recovered frames can closely resemble the ground truth frames. In addition, there is often a large difference between the previews frame and the current frame, and it can also be seen in this visualization that our model generates very reasonable predictions in regions where no reference can be found.

Figure~\ref{fig:rc_cl_dnn_quality} further shows the partial video recovery results. We receive and decode video frames in a WiFi network environment. In this setting, many frames can only be partially decoded and our recovery model can recover these corrupted frames. We fill the decoded part of the frame into the recovered frame for all of the schemes such that the overall video quality is higher than the whole frame prediction. As we can see, our recovery without the binary point code yields 0.6-5dB PSNR improvement and 0.01-0.04 SSIM improvement over reusing the previous frame. The binary point code further increases PSNR by 4-8.5dB and increases SSIM by 0.04-0.06, respectively. 

Moreover, the gap between our video recovery without the binary point code and reusing the previous frame becomes larger because  $I_{part}$ allows the network to get an accurate hint to infer the content of the current frame. Similarly, the gap between the performance of our recovery with the binary point code and the other algorithms increases a lot compared to Figure~\ref{fig:rc_dnn_quality} likely because the model learns a stronger association between RGB frame content and binary point code in the successfully decoded part and better utilizes the learned binary code to generate predictions in the missing part.

\begin{figure*}[h!]
  \centering
  \includegraphics[width=0.8\textwidth]{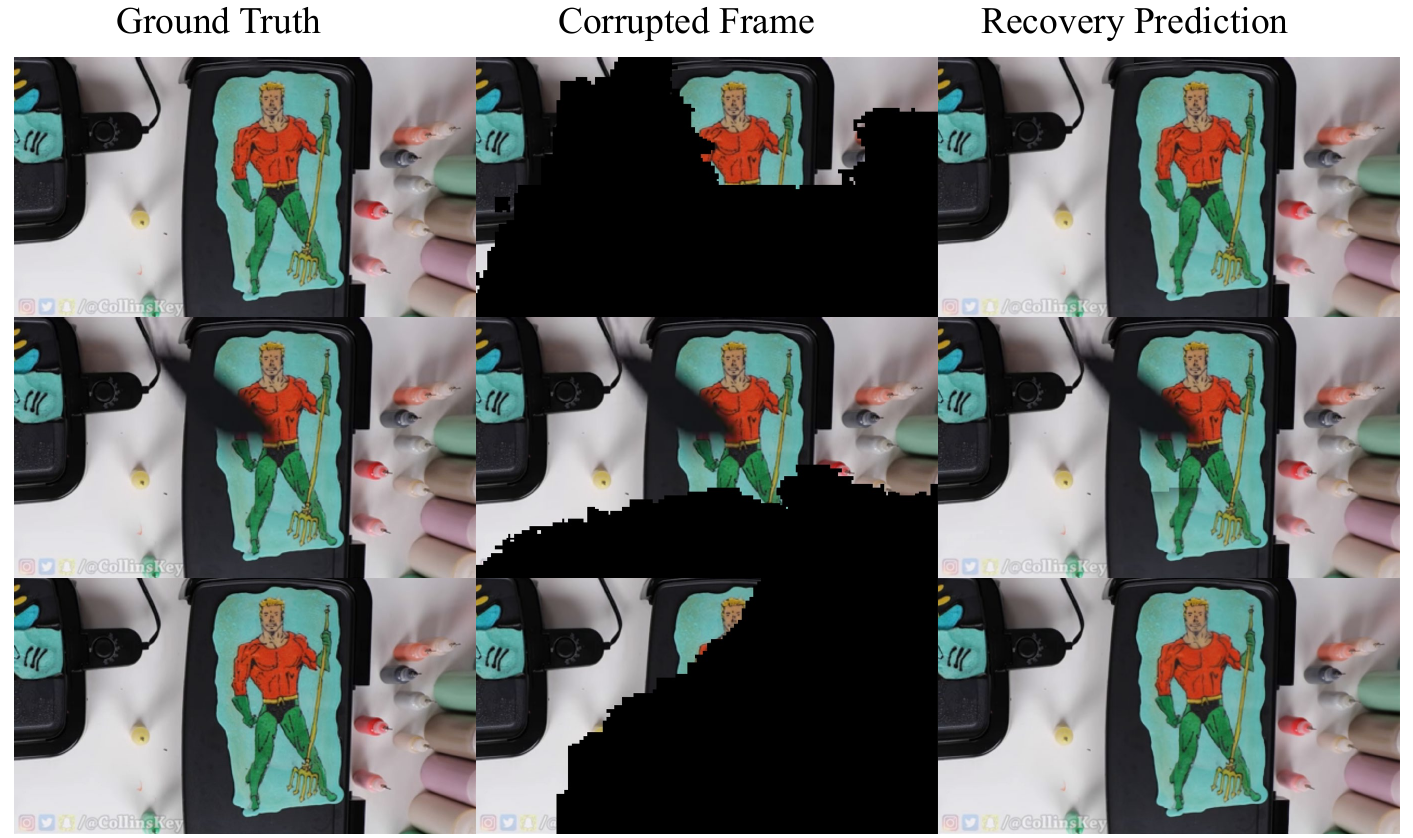}
  \caption{Visualization of video concealment results. Cases are sampled from REDS4~\cite{Nah_2019_CVPR_Workshops_REDS}.}
  \label{fig:vis_conceal}
\end{figure*}

Figure~\ref{fig:vis_conceal} plots the visualization of our error concealment results (\ie, recovery from partially corrupted frames). The recovered frames are also very similar to the original video frames. These results demonstrate the effectiveness of our video recovery for both completely lost or partially corrupted frames. 

\begin{figure}[h!]
  \centering
  \vspace{-10pt}
  \includegraphics[width=0.5\columnwidth]{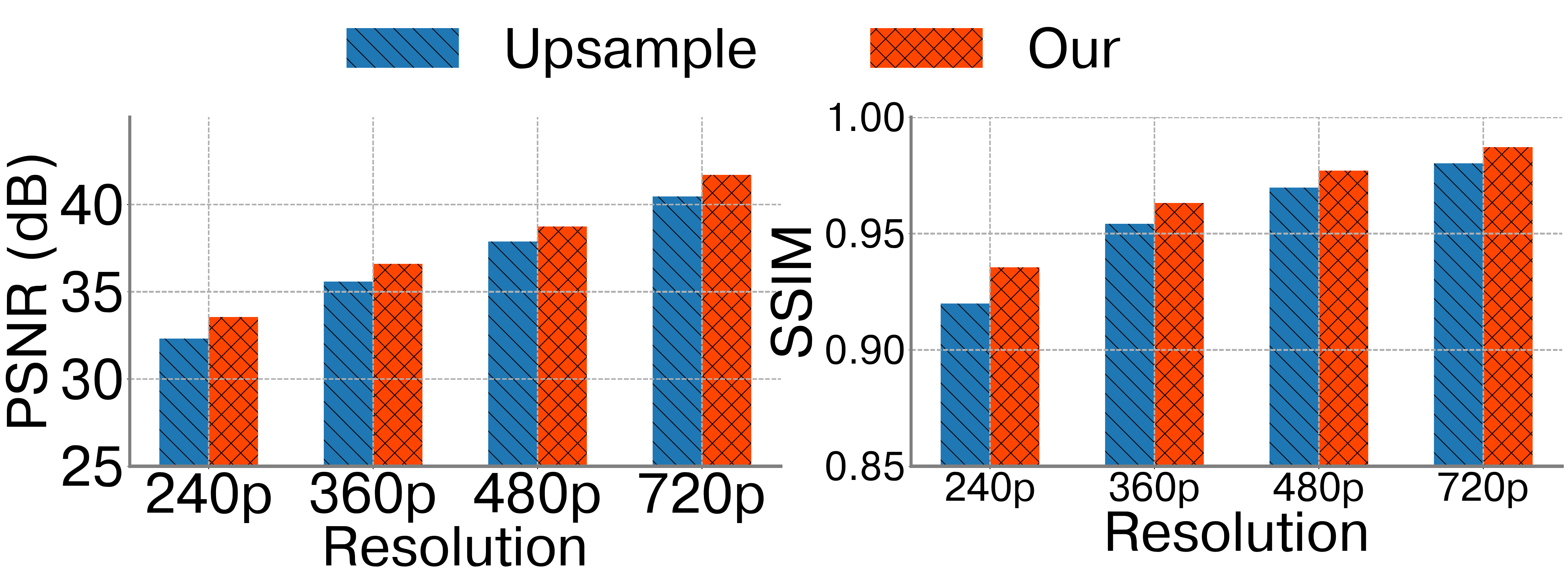}
  \caption{Performance results of video super-resolution.}
  \label{fig:sr_dnn_quality}
\end{figure}

\begin{figure*}[h!]
  \centering
  \includegraphics[width=\textwidth]{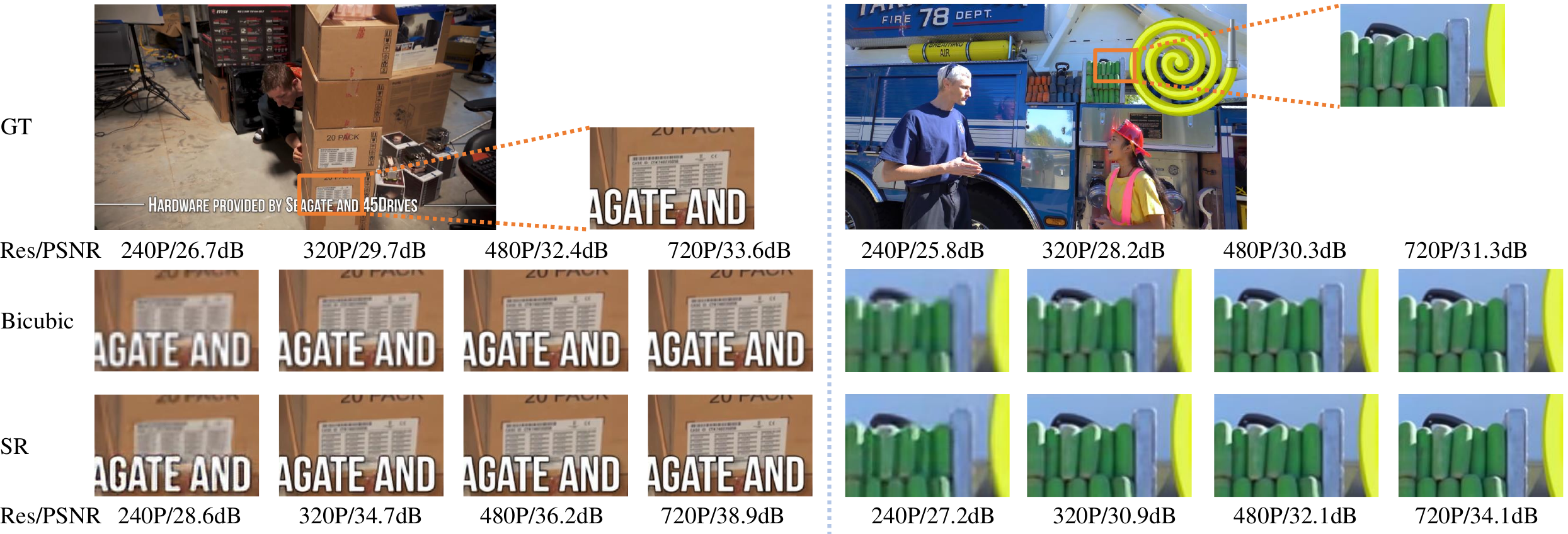}
  \vspace{-5pt}
  \caption{Visualization of video super-resolution results. Two examples are shown to demonstrate that our super-resolutio model achieves significant results at four different scales of video super-resolution.}
  \label{fig:vis_sr}
\end{figure*}

\para{DNN Performance of video super-resolution: } Figure~\ref{fig:sr_dnn_quality} compares the performance of our video super-resolution with upsampling. As we can see, our SR improves the PSNR and SSIM by 1.2dB, 1.1dB, 1dB, and 1.3dB; 0.015, 0.01, 0.007, and 0.008 at 240p, 360p, 480p, and 720p, respectively. The lower resolution video frames yield a higher improvement, as expected. Figure~\ref{fig:vis_sr} plots the visualization of video super-resolution results. Our proposed super-resolution algorithm delivers stable video frame quality improvement at all resolutions. 

\subsection{System Performance}

In this section, we evaluate the system performance in terms of video QoE. Note that we downscale the throughput for all network traces so that their throughput falls into the range between the highest and lowest video bit rates.  
The average downscaled throughput among all the network traces is around 1-2Mbps. 


\para{QoE performance of video recovery: }To evaluate the QoE performance of video recovery, we consider three schemes: (i) without recovery model, (ii) without recovery-aware ABR, and (iii) our approach. Note that (ii) means we still perform video recovery for lost or late frames but select the bitrate without taking into account the benefits and cost of video recovery.  

\begin{figure}[H]  
  \centering  
  \begin{minipage}{0.5\textwidth}  
    \centering
    \includegraphics[width=\columnwidth]{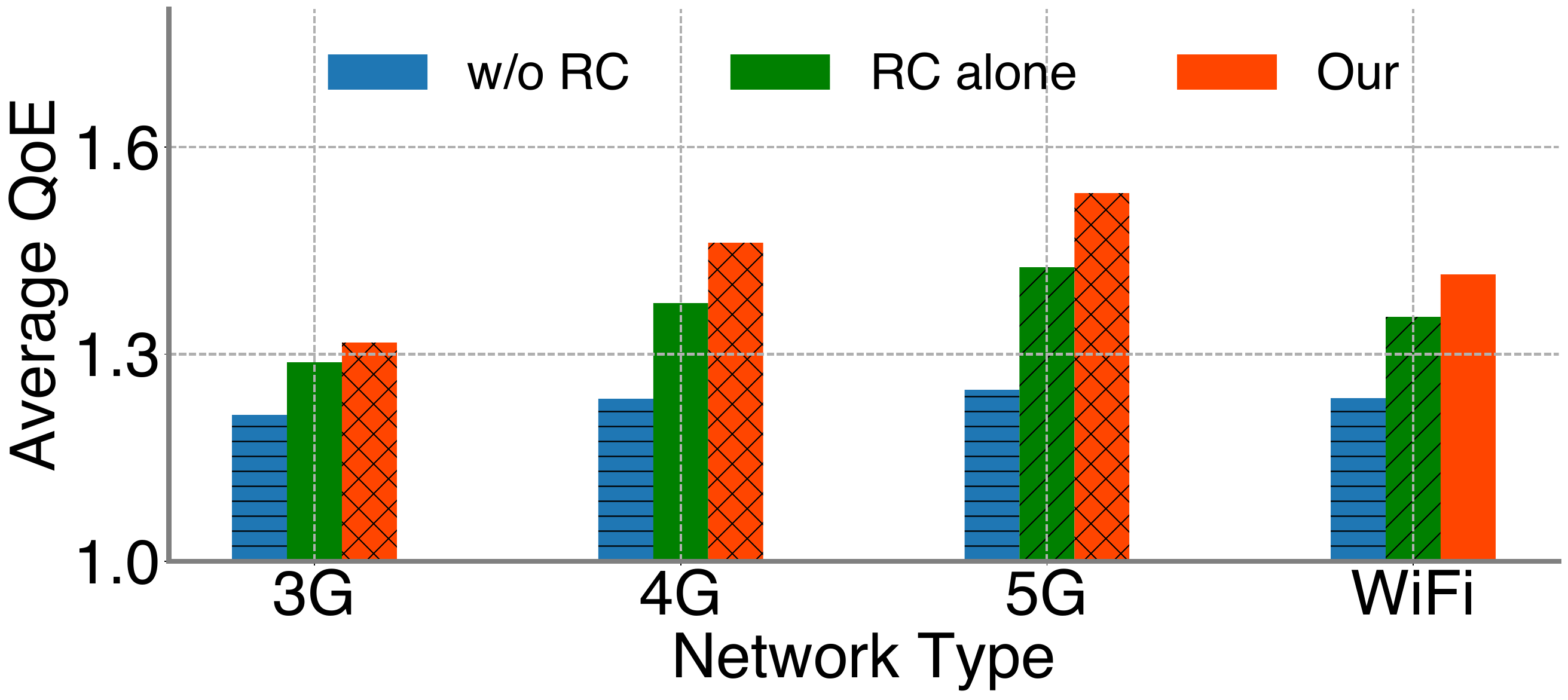}
    \vspace{-10pt}
    \caption{Qoe of Recovery-only schemes across different network traces.}
    \label{fig:rc_only_qoe}
  \end{minipage}%
  \begin{minipage}{0.45\textwidth}  
    \centering  
    \resizebox{0.85\columnwidth}{!}{%
    \begin{tabular}{c|c|c|c|c}
        \toprule
        & 3G & 4G & 5G & WiFi \\
        \midrule
        w/o RC & -0.88 & -9.24 & -11.86 & -1.99 \\
        RC alone & 0.1 & -0.48 & -1.21 & -0.16 \\
        Our & 0.4 & 0.07 & 0.19 & 0.11 \\
        \bottomrule
    \end{tabular}
    }
    \captionsetup{type=table}
    \caption{QoE of recovered frames}
    \label{tab:qoe_rc_frames}
  \end{minipage}  
\end{figure}

Figure~\ref{fig:rc_only_qoe} shows the QoE performance of recovery-only schemes across different network traces. We make the following observations. First, video recovery alone improves the average QoE by 6.3\%, 11.2\%, 14.2\%, and 9.6\% in 3G, 4G, 5G, and WiFi, respectively, because it can recover lost and late frames such that the rebuffering time can be effectively reduced. 

Second, our recovery-aware algorithm improves over without recovery by 8.6\%, 18.3\%, 22.8\%, and 14.5\% in 3G, 4G, 5G, and WiFi, respectively, and improves over recovery alone by 2.2\%, 6.4\%, 7.5\%, and 4.5\% in 3G, 4G, 5G, and WiFi, respectively because it is aware of the usage of recovery for the next frames such that the bitrate can be chosen more wisely to maximize the system QoE. 

Third, comparing the results across different types of networks, we observe 5G enjoys the largest improvement because more frames require video recovery as we will show next.

\begin{figure*}
\centering
\begin{minipage}[ht]{1\columnwidth}
\begin{minipage}{0.5\columnwidth}
    \begin{figure}[H]
    \centering
    \vspace{-10pt}
    \subfigure[Average downscaled throughput.]{%
        \includegraphics[width=0.46\columnwidth]{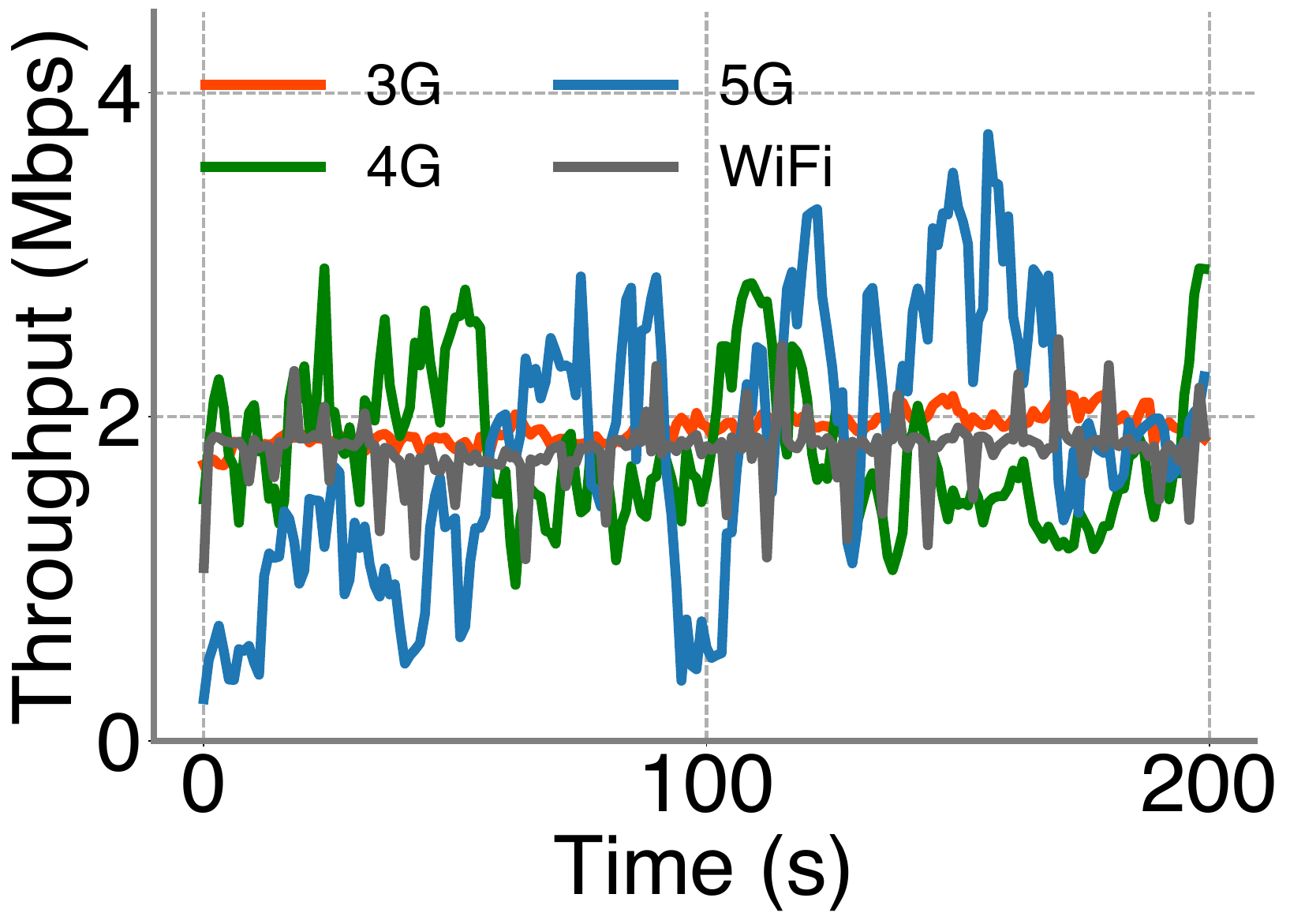}
        \vspace{-15pt}
        \label{fig:throughput}
    }
    \subfigure[Percentage of recovered frames]{%
        \includegraphics[width=0.46\columnwidth]{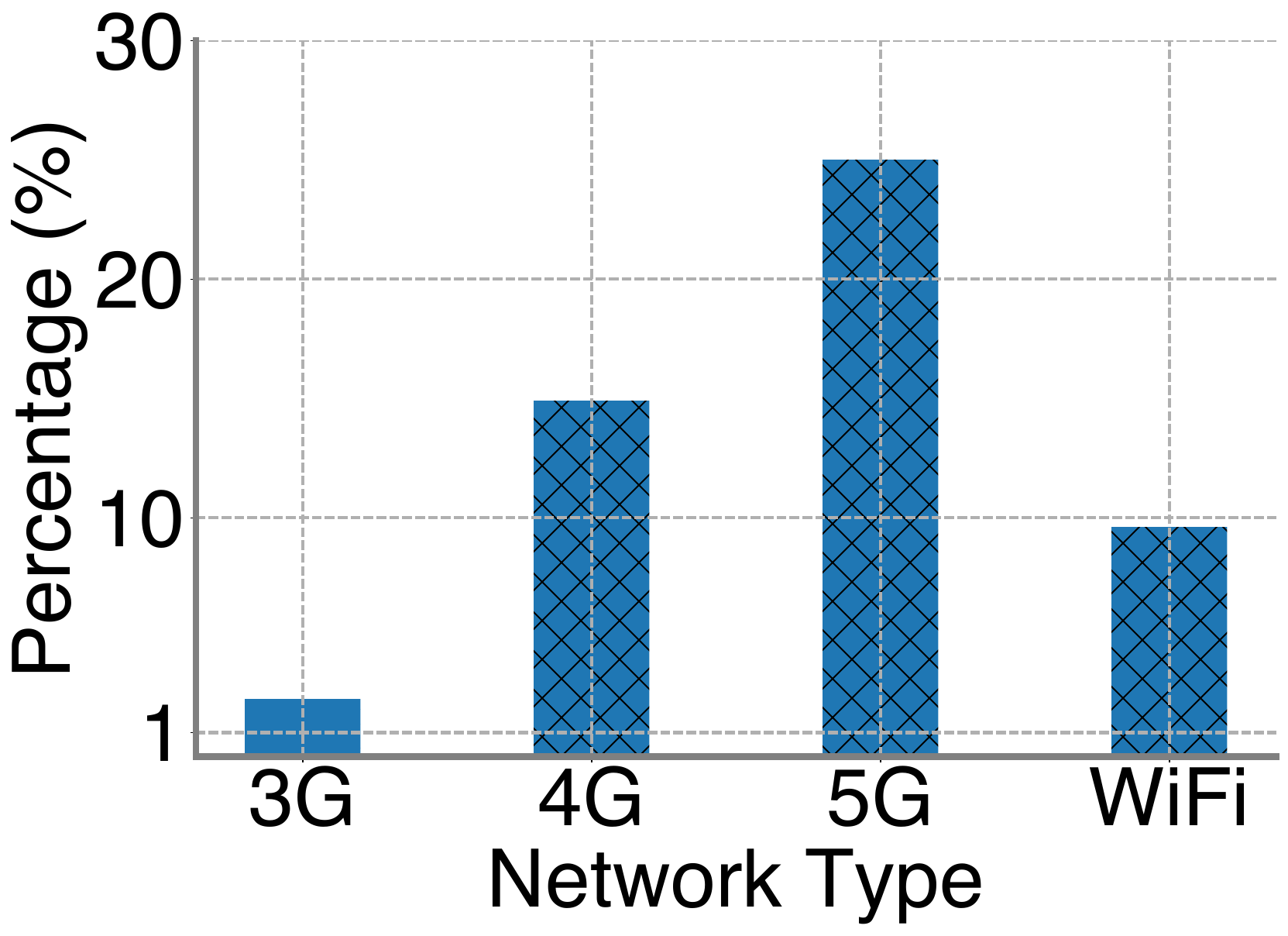}
        \vspace{-15pt}
        \label{fig:rc_percentage}
    }
    \caption{Analysis of the QoE performance of video recovery.}
    \label{fig:analysis_recovery}
    \end{figure}
\end{minipage}
\hspace{5pt}
\begin{minipage}{0.5\columnwidth}
    \begin{figure}[H]
    \vspace{-10pt}
    \subfigure[QoE.]{%
        \includegraphics[width=0.46\columnwidth]{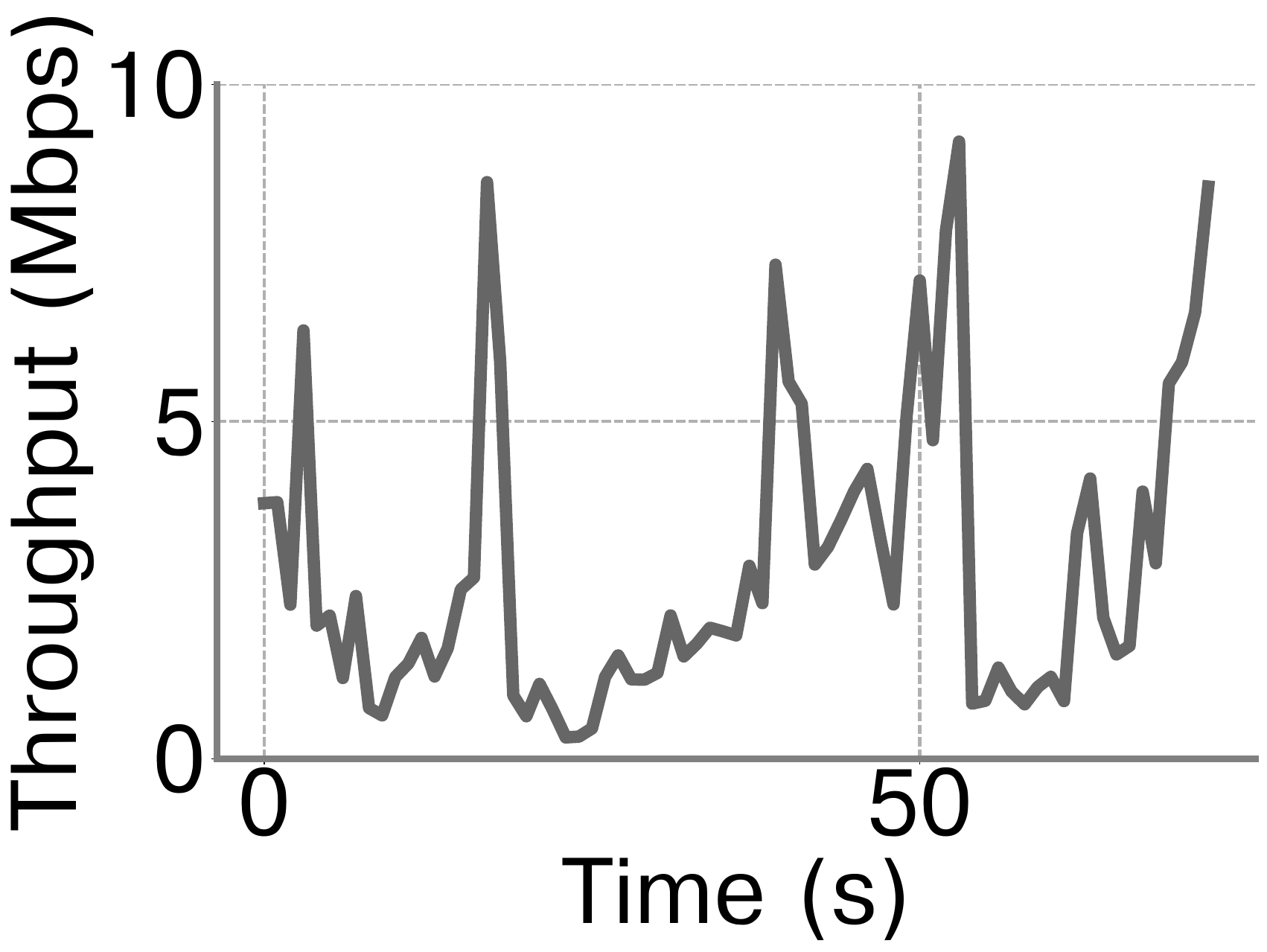}
        \vspace{-15pt}
        \label{fig:sample_thrp}
    }
    \subfigure[Throughput]{%
        \includegraphics[width=0.46\columnwidth]{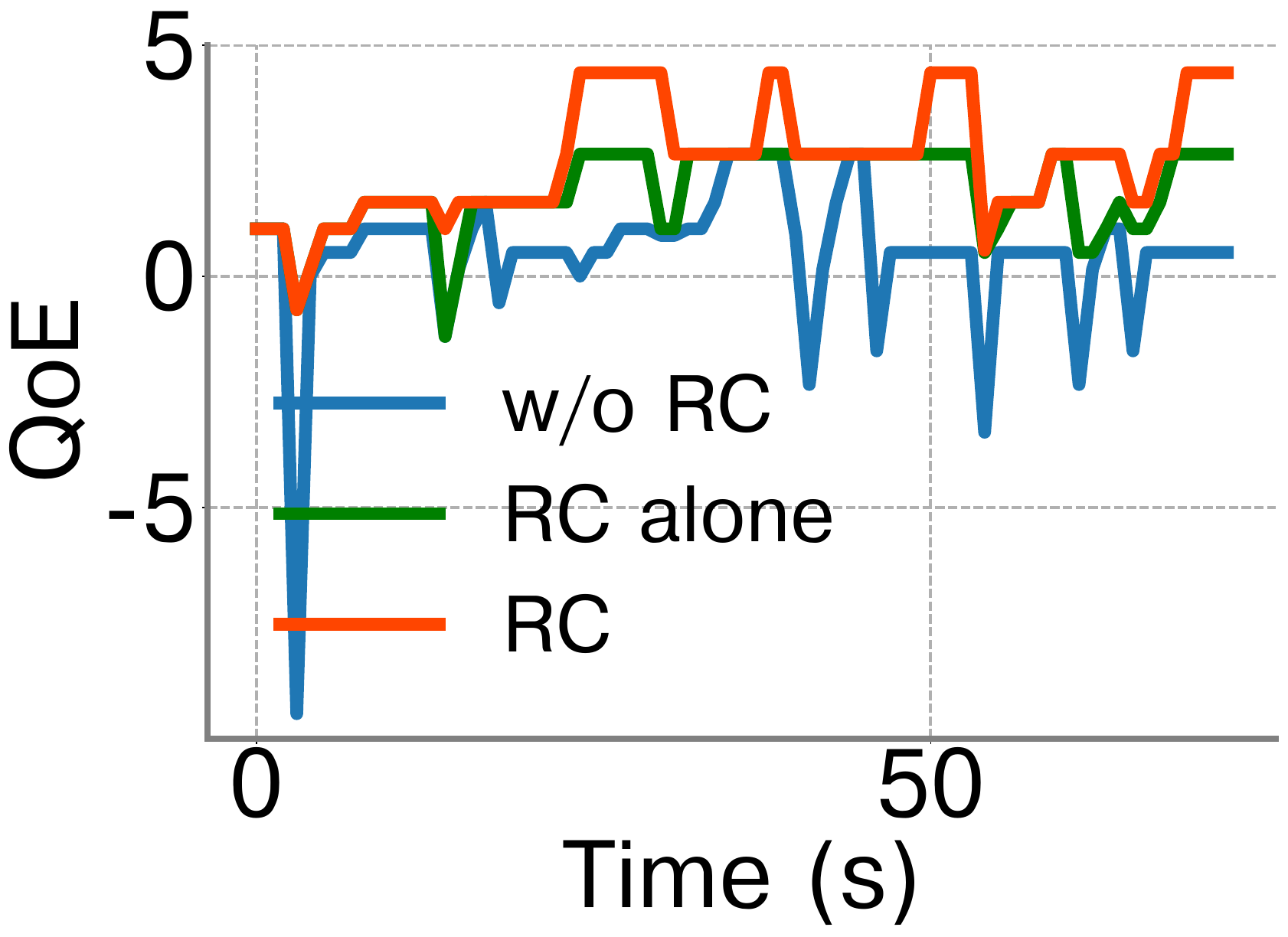}
        \vspace{-15pt}
        \label{fig:sample_qoe}
    }
    \caption{Sample time series of throughput and corresponding QoE under 5G network.}
    \label{fig:sample_thrp_qoe}
\end{figure}
\end{minipage}
\end{minipage}
\end{figure*}

Figure~\ref{fig:throughput} shows the average downscaled throughput of different network traces. We see a large fluctuation in 5G traces. Figure~\ref{fig:rc_percentage} reports the percentage of recovered frames. As 5G has the largest throughput fluctuation, many video frames are not received in time and require video recovery. Meanwhile, even 4G and WiFi see close to 10\% or more video frames that require video recovery. These numbers suggest that video recovery is important due to challenging network conditions. Figure~\ref{fig:sample_thrp} further shows a sample time series of throughput. We find that the scheme without recovery cannot sustain a good QoE when the throughput varies a lot. Recovery alone has a more stable QoE but sometimes gets below 0 due to the rebuffering overhead. Our approach always chooses the bitrate that yields the best QoE. 

Table~\ref{tab:qoe_rc_frames} reports the average QoE of the recovered video frames only. Video recovery alone improves the QoE for the recovered frames by 1.26 - 10.65. The improvement comes mostly from reduced rebuffering time. 
Incorporating recovery-aware ABR further increases the QoE by 0.25 - 1.4. 

\begin{figure*}
\centering
\begin{minipage}[ht]{1\columnwidth}
\begin{minipage}{0.49\columnwidth}
    \begin{figure}[H]
    \centering
    \vspace{-10pt}
    \includegraphics[width=\columnwidth]{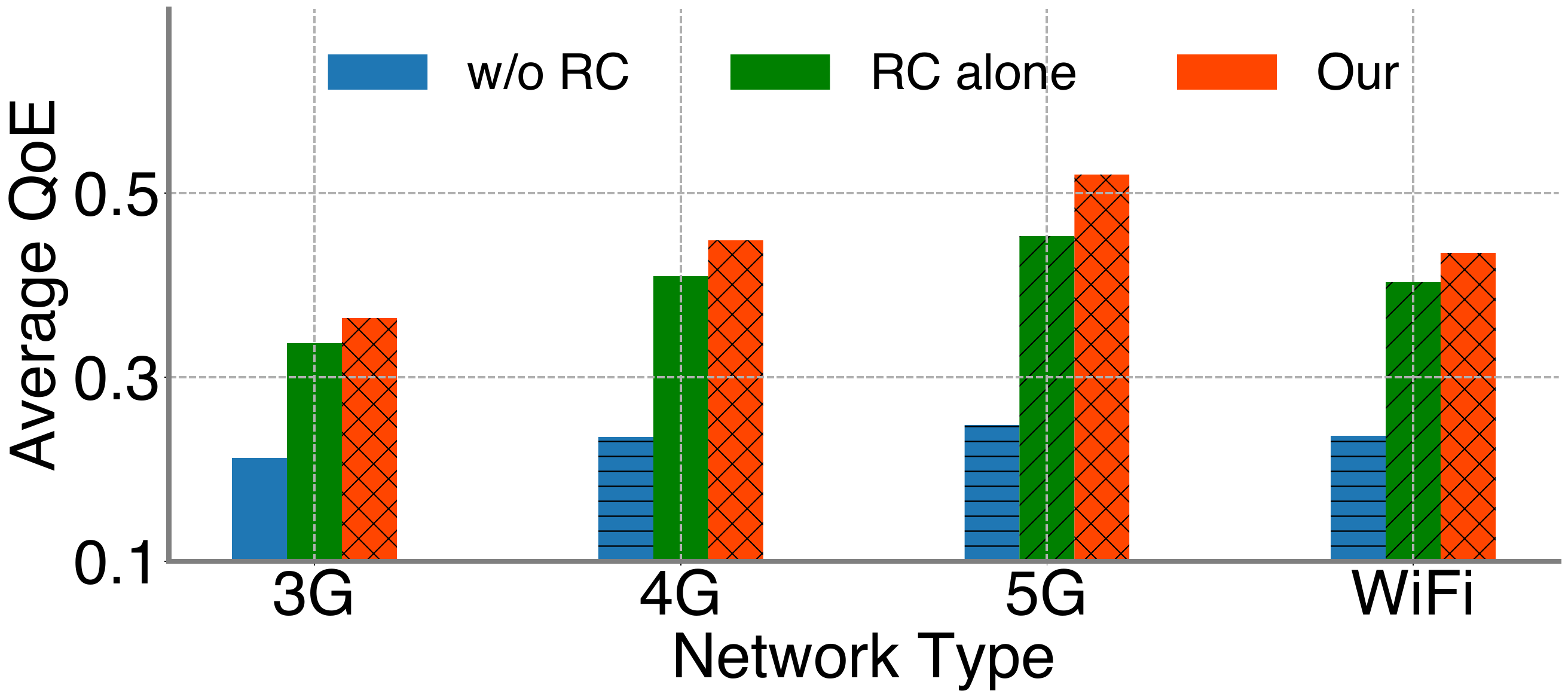}
    \caption{Qoe of Recovery-only schemes across different network traces.}
    \label{fig:rc_lossy_qoe}
    \end{figure}
\end{minipage}
\hspace{5pt}
\begin{minipage}{0.49\columnwidth}
    \begin{figure}[H]
    \centering
    \vspace{-10pt}
    \includegraphics[width=\columnwidth]{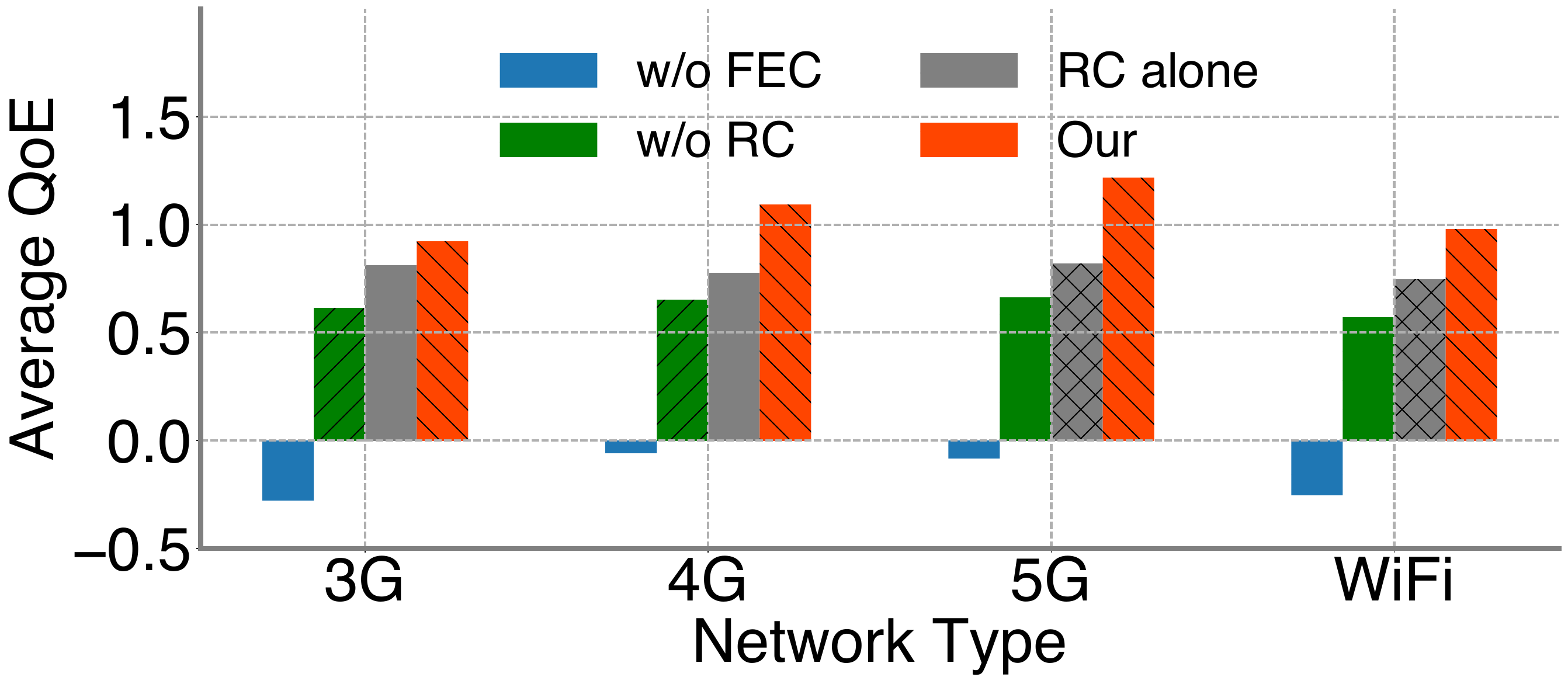}
    \caption{QoE of Recovery-only schemes with and without FEC under different network traces.}
    \label{fig:rc_fec_lossy_qoe}
\end{figure}
\end{minipage}
\end{minipage}
\end{figure*}

\para{QoE performance without FEC under lossy networks: } Figure~\ref{fig:rc_lossy_qoe} shows the QoE performance of recovery-only schemes across different network traces. Under this setting, we do not enable FEC for loss recovery. For (i), we reuse the last frame when a video frame is late or lost. For (ii) and (iii), our recovery model recovers both lost frames and late frames. Under a lossy network environment, we observe that video recovery alone improves the average QoE by 58.9\%, 74.3\%, 82.7\%, and 70.6\% in 3G, 4G, 5G, and WiFi, respectively. Our approach improves over that without recovery by 71.8\%, 90.8\%, 110\%, and 84.3\% in 3G, 4G, 5G, and WiFi, respectively, and improves over recovery alone by 8.1\%, 9.5\%, 14.6\%, and 8\% in 3G, 4G, 5G, and WiFi, respectively. We find that the improvement of our approach over baselines increases a lot under the lossy network environment because reusing the previous frames is not effective under many consecutive lost/late frames (as shown in Figure~\ref{fig:rc_dnn_quality}), which is more likely under a lossy environment. However, our recovery model can recover many frames with little degradation so its QoE performance is much better. 

\para{QoE performance with FEC under lossy networks: } So far, we disable FEC in our algorithm. Next we further jointly optimize FEC and video recovery. Figure~\ref{fig:rc_fec_lossy_qoe} compares our algorithm but disable FEC (w/o FEC) with all other schemes with FEC, where the amount of FEC is determined based on our lookup table. We offline build separate lookup tables that map the packet loss rates to desired FEC levels for different schemes. Our joint optimization yields 51\%, 68\%, 83\%, and 72\% improvement over no recovery in 3G, 4G, 5G, and WiFi, respectively. The corresponding improvements over recovery alone are 13\%, 41\%, 48\%, and 31\%, respectively. Also, it outperforms no FEC by 1.2, 1.15, 1.3, and 1.23 in QoE, respectively. These results show that (i) FEC plays an important role under lossy network conditions, (ii) the desired amount of FEC depends on the recovery and ABR algorithms, and (iii) each component in our recovery model (\ie, recovery alone, recovery-aware, and joint optimization of FEC and recovery) is beneficial. 



\begin{figure*}
\centering
\begin{minipage}[ht]{1\columnwidth}
\begin{minipage}{0.49\columnwidth}
    \begin{figure}[H]
    \centering
    \vspace{-10pt}
    \includegraphics[width=\columnwidth]{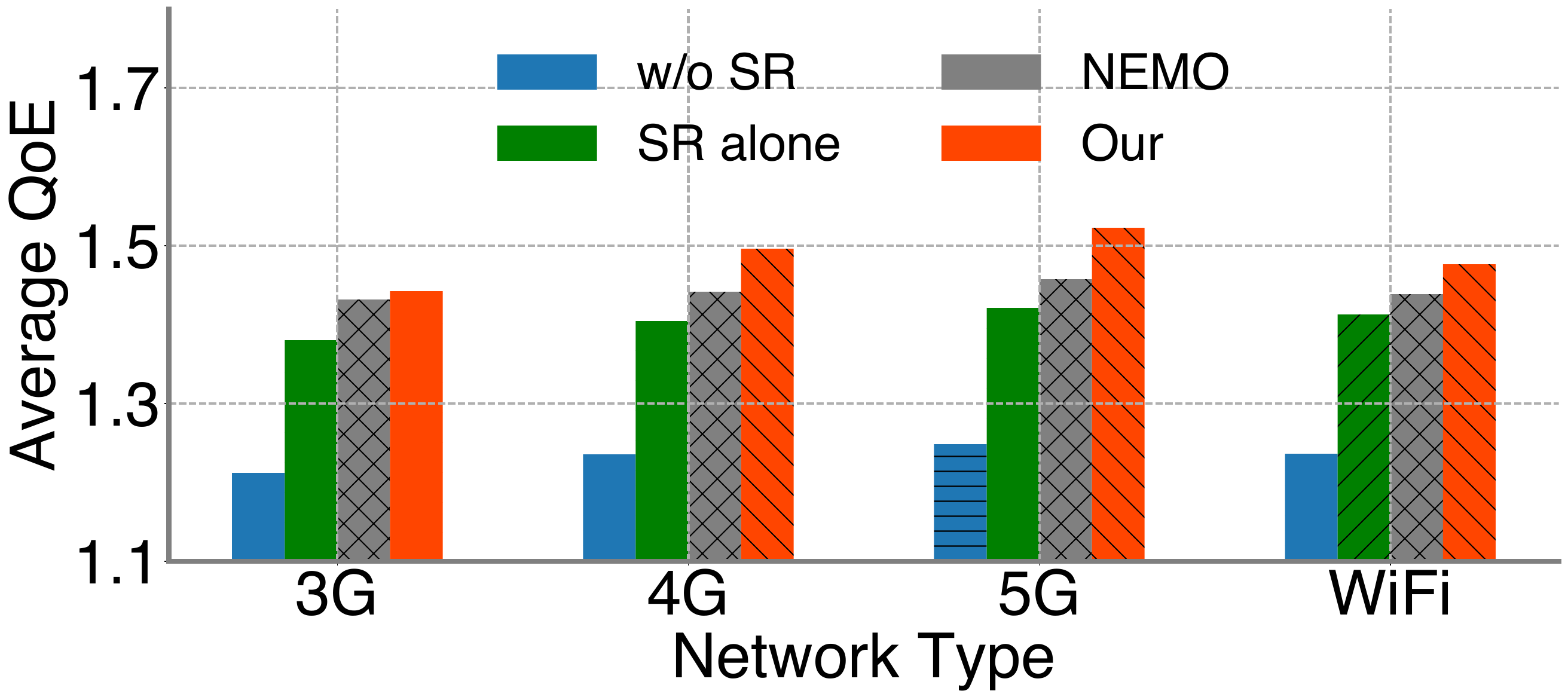}
    \caption{Qoe of SR-only schemes across different network traces.}
    \label{fig:sr_only_qoe}
    \end{figure}
\end{minipage}
\hspace{5pt}
\begin{minipage}{0.49\columnwidth}
    \begin{figure}[H]
    \centering
    \vspace{-10pt}
    \includegraphics[width=\columnwidth]{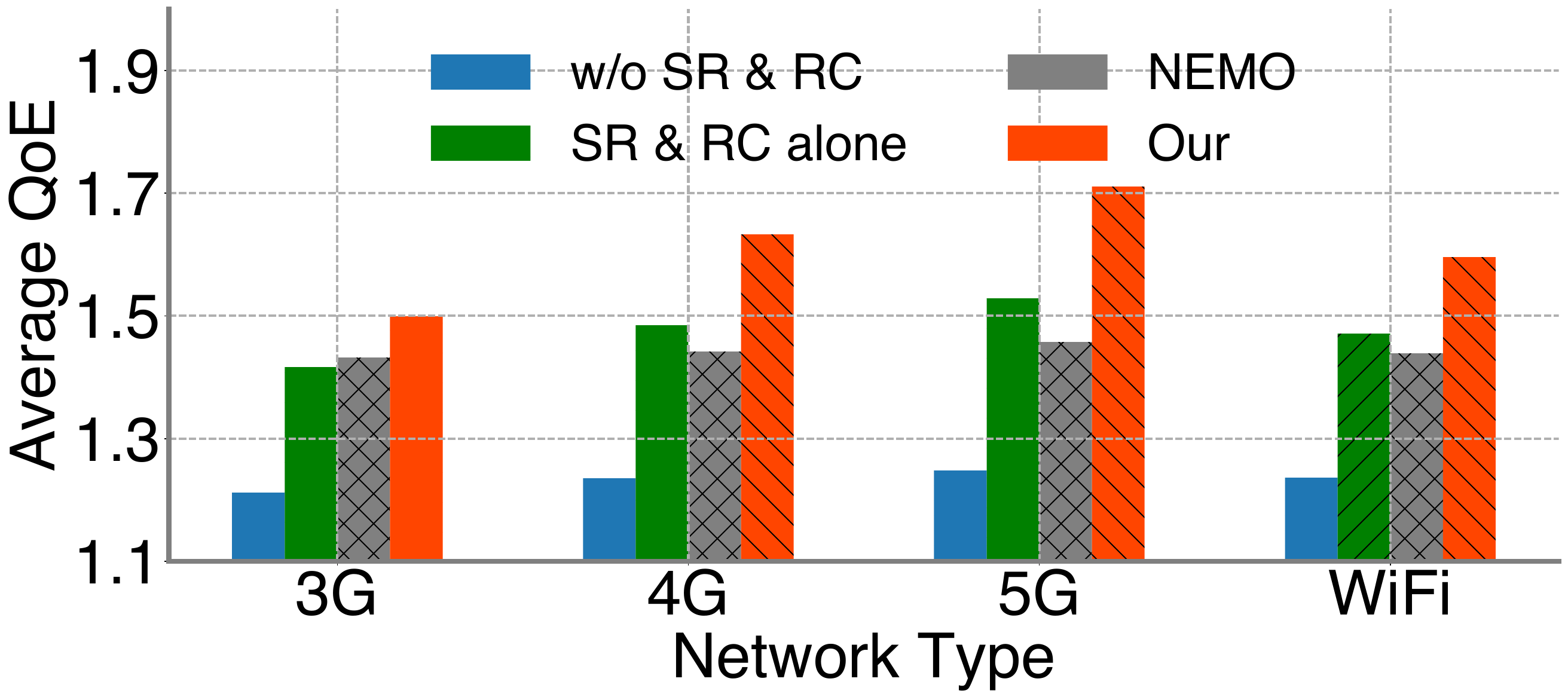}
    \caption{Qoe of Recovery \& SR schemes across different network traces.}
    \label{fig:sr_rc_qoe}
\end{figure}
\end{minipage}
\end{minipage}
\end{figure*}

\para{QoE performance of video super-resolution: }Figure~\ref{fig:sr_only_qoe} compares the QoE of our super-resolution with (i) without SR, (ii) SR alone using our model, and (iii) NEMO~\cite{yeo2020nemo}.  Our SR-aware approach significantly outperforms all the other algorithms. Its improvement over (i), (ii), and (iii) are 18\%, 21\%, 22\%, and 19\%; 4.5\%, 6.5\%, 7.1\%, and 4.5\%; 0.7\%, 3.8\%, 4.5\%, and 2.7\% in 3G, 4G, 5G, and WiFi, respectively. SR alone brings 12\%-14\% improvement, and SR-aware ABR further brings 4\%-7\% improvement, which shows the importance of joint design of ABR and SR. 

\para{QoE performance of video recovery and super-resolution: } Figure~\ref{fig:sr_rc_qoe} compares the QoE of (i) without SR or recovery, (ii) SR and recovery alone, (iii) NEMO, and (iv) our final algorithm. Our algorithm out-performs (i), (ii), and (iii) by 23.7\%, 32.2\%, 37.1\%, and 29\%; 5.9\%, 10\%, 11.9\%, and 8.4\%; 4.7\%, 13.2\%, 17.4\%, and 10.9\% in 3G, 4G, 5G, and WiFi, respectively. It can be found that both SR and Recovery play a significant effect, and combined with our enhancement aware ABR strategy, our method achieves the best performance. It out-performs NEMO by 4.7\%-17.4\%, and even SR and Recovery alone out-perform NEMO in all cases except 3G, because NEMO does not have recovery and has to reuse the previous frames for late or lost video frames. 3G is better for NEMO due to fewer lost/late video frames.

\zhaoyuan{
\subsection{System Latency and Resource Usage}

\para{System latency: } At the start of video streaming, we establish TCP and QUIC transmission sessions. The binary code, with a size of 1KB, can be encapsulated into a single TCP packet. Consequently, the TCP latency for each frame is expected to be approximately equivalent to the round-trip time (RTT). Given that the decoding and model inference processes of a frame can occur simultaneously with the receiving process of subsequent frames, the total latency can be viewed as the sum of the decoding time and the duration required for neural enhancement or recovery. The decoding time of 240p, 360p, 480p, 720p, and 1080p videos is 1.8, 2.3, 2.9, 4.1, and 6.2ms on the iPhone 12, respectively. Our model adds an additional 22ms for both enhancement and recovery, regardless of the video resolution. This results in a total latency of under 33 ms, demonstrating real-time processing capability in our system.

\para{CPU usage and energy consumption: } We also measure the CPU utilization and energy consumption with and without our model. We only evaluate the neural video recovery because both video recovery and enhancement share a similar model structure and exhibit identical inference time. This similarity implies comparable CPU usage and energy consumption between the two models. Without DNN processing, iPhone 12's CPU utilization is 28\% and the energy consumption is 0.04J per frame. Under 20\% frame losses, the corresponding numbers are 37\% and 0.05J, and under 100\% frame losses, they are 68\% and 0.07J. Consequently, with each frame undergoing neural recovery or enhancement, the expected battery life decreases from 13.2 hours to 7.5 hours.

}

\section{Conclusion}
\label{sec:conclusion}

As mobile video streaming becomes increasingly popular, we develop video recovery and SR for mobile clients. We further show it is important to adapt the video bit rate according to the receiver enhancement. We develop and implement our video streaming system. Our extensive evaluation shows that our approach yields significant improvement in 3G, 4G, WiFi, and 5G networks. 

\bibliographystyle{abbrv}
\bibliography{paper,video}


\end{document}